\documentclass[a4paper,11pt]{article}
\usepackage{jheppub}

\usepackage[utf8]{inputenc}
\usepackage{amsmath}
\usepackage{amssymb}
\usepackage{mathrsfs}
\usepackage{graphicx}
\usepackage{bbold}
\usepackage{comment}
\usepackage[dvipsnames]{xcolor}
\usepackage{braket}
\usepackage{bm,bbm}
\usepackage{physics}
\usepackage{tikz}

\newcommand\beq{ \begin{eqnarray} }
\newcommand\eeq{ \end{eqnarray} }
\newcommand{\SU}{\mathrm{SU}}

\preprint{YITP-24-79, RIKEN-iTHEMS-Report-24}

\title{\boldmath DMRG study of the theta-dependent mass spectrum in the $2$-flavor Schwinger model}

\author[a,b]{Etsuko Itou,}
\author[a,b]{Akira Matsumoto}
\author[a]{and Yuya Tanizaki}

\affiliation[a]{
Yukawa Institute for Theoretical Physics, Kyoto University,\\
Kitashirakawa Oiwakecho, Sakyo-ku, Kyoto 606-8502 Japan}

\affiliation[b]{
Interdisciplinary Theoretical and Mathematical Sciences Program (iTHEMS), RIKEN,\\
2-1 Hirosawa, Wako, Saitama 351-0198 Japan}

\emailAdd{itou(at)yukawa.kyoto-u.ac.jp}
\emailAdd{akira.matsumoto(at)yukawa.kyoto-u.ac.jp}
\emailAdd{yuya.tanizaki(at)yukawa.kyoto-u.ac.jp}

\abstract{
We study the $\theta$-dependent mass spectrum of the massive $2$-flavor Schwinger model in the Hamiltonian formalism 
using the density-matrix renormalization group(DMRG).
The masses of the composite particles, the pion and sigma meson, are computed by two independent methods.
One is the improved one-point-function scheme, 
where we measure the local meson operator coupled to the boundary state and extract the mass from its exponential decay.
Since the $\theta$ term causes a nontrivial operator mixing, 
we unravel it by diagonalizing the correlation matrix to define the meson operator.
The other is the dispersion-relation scheme, a heuristic approach specific to Hamiltonian formalism.
We obtain the dispersion relation directly by measuring the energy and momentum of the excited states.
The sign problem is circumvented in these methods, and their results agree with each other even for large $\theta$.
We reveal that the $\theta$-dependence of the pion mass at $m/g=0.1$ 
is consistent with the prediction by the bosonized model.
We also find that the mass of the sigma meson satisfies the semi-classical formula, 
$M_{\sigma}/M_{\pi}=\sqrt{3}$, for almost all region of $\theta$.
While the sigma meson is a stable particle thanks to this relation, 
the eta meson is no longer protected by the $G$-parity and becomes unstable for $\theta\neq 0$.
}

\begin{document}
\maketitle
\flushbottom

\section{Introduction}
\label{sec:introduction}

Lattice gauge theories give the foundation of strongly coupled quantum field theories (QFTs). 
The Monte Carlo simulation based on its Euclidean path integral has uncovered various non-perturbative aspects, 
including properties of hadrons in quantum chromodynamics (QCD), 
and those results are successfully comparable to experimental results~\cite{FlavourLatticeAveragingGroupFLAG:2021npn}.
One of the notable achievements in the Lattice Monte Carlo study is the precise prediction of the hadron mass spectra with only a few input parameters. 
Indeed, the mass spectra of the various hadrons cannot be explained by the sum of the renormalized quark masses of the constituents, and the contributions from interactions need to be evaluated non-perturbatively.
However, it is difficult to perform the Monte Carlo simulation when it suffers from the sign problem, 
which typically appears in the non-zero chemical potential or with the topological $\theta$ term in QCD~\cite{Nagata:2021ugx}. 
We need to expand the scope of numerical studies of QFTs to uncover the new physics hidden in such regions. 

Quantum computing and tensor network methods based on the Hamiltonian formalism of lattice gauge theories~\cite{Kogut:1974ag} 
have recently gotten attention as an alternative numerical approach. 
These methods directly approximate wave functionals of low-energy states, 
so they are based on a completely different criterion from that of the importance sampling. 
The development of these complementary frameworks may enable us to obtain new information that is difficult to extract 
from the Euclidean path integral.

With the current of the times, 
we have developed three distinct methods for computing mass spectra of lattice Hamiltonian gauge theories in ref.~\cite{Itou:2023img}:
(1)~The correlation-function scheme extracts the particle mass by measuring the spatial correlator, 
which is similar to the conventional Euclidean method.
(2)~The one-point-function scheme utilizes the boundary effect as a source of the composite particles, 
and the mass is computed from the exponential decay of the one-point function.
(3)~The dispersion-relation scheme directly computes the dispersion relation by generating the low-lying states at finite volume, 
where the states of each meson with the momentum excitation can be identified via the quantum numbers.
The last scheme, in particular, is specific to the Hamiltonian formalism, unlike the other two.
In ref.~\cite{Itou:2023img}, we applied these methods to the $2$-flavor massive Shcwinger model at $\theta=0$ 
using the density-matrix renormalization group (DRMG)~\cite{White:1992zz,White:1993zza,Schollw_ck_2005,Schollw_ck_2011}, 
and we confirmed that the mass spectra obtained by these methods are almost consistent with each other and also with theoretical predictions. 

The Schwinger model~\cite{Schwinger:1962tp} is the $(1+1)$d quantum electrodynamics, showing charge confinement as 4d QCD. 
Its Euclidean formulation suffers from the sign problem at nonzero $\theta$ angles, but many of its nonperturbative properties 
are analytically calculable in the massless limit~\cite{Lowenstein:1971fc, Casher:1974vf, Coleman:1975pw, 
Manton:1985jm, Hetrick:1988yg, Jayewardena:1988td, Sachs:1991en, Adam:1993fc, 
Hetrick:1995wq, Narayanan:2012du, Narayanan:2012qf, Lohmayer:2013eka, Tanizaki:2016xcu}. 
Numerically, the $\theta$-dependence of the pion mass for the $2$-flavor case is computed 
using the reweighting technique for the conventional importance sampling method, 
which gives reliable results up to $\theta/2\pi \lesssim 0.25$ and confirms some analytic predictions~\cite{Fukaya:2003ph}. 
However, the sign problem becomes severe beyond that point and we need further different approaches 
to numerically study the physics at larger $\theta$.
Hence, the Schwinger model provides an interesting testing ground for new computational methods, 
and indeed there have been many studies on the model with nonzero $\theta$ by the tensor network or quantum algorithm~\cite{Banuls:2013jaa, Banuls:2015sta, Banuls:2016lkq, 
Buyens:2015tea, Buyens:2016ecr, Buyens:2016hhu, Buyens:2017crb, Funcke:2019zna, Chakraborty:2020uhf, 
Kharzeev:2020kgc, Honda:2021aum, Thompson:2021eze, Honda:2021ovk, Honda:2022edn, Nagano:2023uaq, 
Funcke:2023lli, Angelides:2023bme, Ikeda:2023zil, Dempsey:2023gib, Pedersen:2023asd, Schmoll:2023eez, 
Angelides:2023noe, Ghim:2024pxe, Popov:2024ysn, Kaikov:2024acw}.\footnote{
As another Hamiltonian-based approach, there is a numerical study with the light-cone quantization~\cite{Harada:1993va}. 
The tensor network in the Lagrangian formalism also provides a way to be liberated 
from importance sampling~\cite{Shimizu:2014uva, Shimizu:2014fsa, Butt:2019uul, Yosprakob:2023tyr, Az-zahra:2024gqr}.
For a specific class of $(1+1)$d gauge theories including the Schwinger model, the dual variable 
formulation~\cite{Gattringer:2015nea, Gattringer:2015baa, Gattringer:2018dlw, Sulejmanpasic:2019ytl, Sulejmanpasic:2020lyq} 
or the bosonization~\cite{Ohata:2023sqc, Ohata:2023gru} can be used to eliminate the sign problem in the Monte Carlo methods. 
Among these options, we focus on the Hamiltonian formalism written by the spin variables to fit the quantum computations and tensor networks.}

In this paper, we extend our previous study~\cite{Itou:2023img} 
to investigate the physics of the $2$-flavor Schwinger model in the $\theta\neq 0$ regime. 
At $\theta=0$, the mesons can be characterized by their quantum numbers $J^{PG}$, 
where $J$, $P$, and $G$ are the isospin, parity, and $G$-parity, respectively. 
There are three types of stable mesons; 
pion $\pi_a$ ($J^{PG}=1^{-+}$), sigma meson $\sigma$ ($J^{PG}=0^{++}$), and eta meson $\eta$ ($J^{PG}=0^{--}$).
The low-energy spectrum has been studied analytically using the bosonization technique~\cite{Coleman:1976uz}.
Furthermore, the WKB-type approximation predicts that the sigma-meson mass has the specific ratio with the pion mass, 
$M_\sigma/M_\pi=\sqrt{3}$, and these are confirmed in our previous work~\cite{Itou:2023img}. 
On the other hand, at $\theta \neq 0$, the situation will be changed: 
First, the parity and $G$-parity are no longer good quantum numbers at $\theta \neq 0$, 
and the symmetries can no longer distinguish $\sigma$ and $\eta$. 
This causes the $\sigma$-$\eta$ mixing and the decay channel $\eta\to \pi\pi$ also opens.
Second, the $\theta$-dependence of the pion mass is predicted to behave as 
$M_{\pi}(\theta)\sim g|(m/g)\cos(\theta/2)|^{2/3}$ when the fermion mass is small~\cite{Coleman:1976uz}, 
and the mass of the sigma meson is also predicted to satisfy the relation $M_{\sigma}(\theta)/M_{\pi}(\theta)=\sqrt{3}$, 
but it has not yet been confirmed by the first-principle calculations because of the sign problem. 
Moreover, substituting $\theta=\pi$ into these formulas, the spectrum is expected to become massless. 
Indeed, it is suggested that the massive $2$-flavor Schwinger model at $\theta=\pi$ shows almost conformal behavior 
and is described by the $SU(2)_1$ Wess-Zumino-Witten (WZW) model 
with the marginally relevant $J_L J_R$ deformation~\cite{Coleman:1976uz, Dempsey:2023gib}.
From these theoretical predictions, the extension of the calculation methods from $\theta=0$ to $\theta \neq 0$ is somewhat non-trivial, 
even in the Hamiltonian formalism.

We performed the DMRG calculations and obtained the mass spectra mainly by the two types of calculation methods, 
namely the one-point-function scheme and the dispersion-relation scheme with several improvements to the ones originally proposed 
in ref.~\cite{Itou:2023img}. 
We found that we can precisely calculate the mass spectra of stable mesons, 
and the results of these two schemes are consistent with each other even in the $\theta \neq 0$ regime.
Furthermore, our results indicate that the $\theta$-dependence of the pion mass agrees with the analytic result, 
$|\cos(\theta/2)|^{2/3}$, and the sigma meson is $\sqrt{3}$ times heavier than the pion as predicted.
We also confirmed the nearly conformal behavior at $\theta=\pi$ 
by comparing the one-point functions of the conformal theory on a finite interval with its analytic calculation. 

This paper is organized as follows.
In section~\ref{sec:theory}, 
we revisit the continuum $2$-flavor Schwinger model and its bosonization focusing on the low-energy mass spectrum.
We review the three methods for computing the mass spectrum in section~\ref{sec:strategy} 
and discuss the extension to $\theta\neq 0$ in section~\ref{sec:strategy_for_theta}.
In section~\ref{sec:result}, 
we present our simulation results of the mass spectra by the one-point and dispersion-relation schemes.
In section~\ref{sec:Result_CFT}, 
we show the one-point functions at $\theta=\pi$ and compare them with the analytic results.
Section~\ref{sec:conclusion} is devoted to the conclusion and discussion.
In appendix~\ref{sec:compact_boson_interval}, 
we explain the analytic calculation of the one-point function in the WZW model on a finite interval.
Appendix~\ref{sec:operators} shows the definition of the observables used in the analysis.
In appendix~\ref{sec:cf_scheme}, 
we summarize the results of the correlation-function scheme for reference.
In appendix~\ref{sec:1pt_func_original}, 
we show the result of the one-point-function scheme with a different setup from the main part.
In appendix~\ref{sec:fate_of_eta}, 
we discuss the behavior of the one-point and correlation functions of the unstable eta meson.

\section{Review of the 2-flavor Schwinger model with the \texorpdfstring{$\theta$}{theta} term}
\label{sec:theory}

In this section, we review the basic properties of the $2$-flavor Schwinger model~\footnote{
For the original reference, see ref.~\cite{Coleman:1976uz}, and section~2 of our previous paper~\cite{Itou:2023img} would also be useful.}.
The Lagrangian of the massive $2$-flavor Schwinger model with the Minkowski metric $\eta_{\mu\nu}=\mathrm{diag}(1,-1)$ is given by 
\begin{equation}
    \mathcal{L}=-\frac{1}{4g^{2}}F_{\mu\nu}F^{\mu\nu}+\frac{\theta}{4\pi}\epsilon_{\mu\nu}F^{\mu\nu}
    +\sum_{f=1}^{2}\left[i\bar{\psi}_{f}\gamma^{\mu}\left(\partial_{\mu}+iA_{\mu}\right)\psi_{f}-m\bar{\psi}_{f}\psi_{f}\right],
    \label{eq:Lagrangian}
\end{equation}
where $F_{\mu\nu}=\partial_{\mu}A_{\nu}-\partial_{\nu}A_{\mu}$ is the field strength, $g$ is the gauge coupling, 
$\theta$ is the vacuum angle describing the background electric flux, 
$m$ is the flavor-symmetric fermion mass, and the index $f$ labels the flavor. 
We note that the mass dimension is $[g]=[m]=1$, and we focus on the strongly-coupled regime, $0<m\ll g$.

\subsection{Composite particles and global symmetry at \texorpdfstring{$\theta=0$}{theta=0}}

The Coulomb interaction in $(1+1)$d is the linear potential.
Therefore, the massive multi-flavor Schwinger model is an example of confining gauge theories, 
and its particle spectrum for $0<m\ll g$ is analogous to the meson spectrum of $(3+1)$d QCD. 

Let us first summarize the physics at $\theta=0$. 
For nonzero fermion mass $m>0$, the $2$-flavor Schwinger model has the mass gap and the asymptotic particle states are well defined. 
The global symmetry provides the quantum numbers for their classification, and the global symmetry at $\theta=0$ is given by 
\begin{equation}
    \frac{\SU(2)_V}{\mathbb{Z}_2}\times (\mathbb{Z}_2)_G. 
    \label{eq:symmetry_theta0}
\end{equation}
Here, $\SU(2)_V/\mathbb{Z}_2$ is the isospin symmetry, which acts on the fermions as a vector-like symmetry, 
$\psi \to \exp(i \varepsilon_a\tau_a)\psi$ and $\bar{\psi}\to \bar{\psi}\exp(-i\varepsilon_a\tau_a)$, and we denote its generators as $J_a$,
\begin{equation}
    J_{a}=\frac{1}{2}\int dx\,\bar{\psi}\gamma^{0}\tau_{a}\psi.
    \label{eq:J_cont}
\end{equation}
The $G$-parity $(\mathbb{Z}_2)_G$ is defined by the combination of the charge conjugation and the $\pi$ rotation in the isospin space, 
\begin{equation}
    G=C e^{i \pi J_y}. 
\end{equation}
In addition to these internal symmetries, we also have the parity quantum number $P$. 

When $\theta=0$, there are three types of the stable particles~\cite{Coleman:1976uz}. 
By analogy to QCD, we call them pion, sigma meson, and eta meson, 
and they correspond to the lowest energy states for the following composite operators: 
\begin{alignat}{3}
 & \pi_{a}|_{\theta=0} &  & =-i\bar{\psi}\gamma^{5}\tau_{a}\psi\quad &  & (J^{PG}=1^{-+}),\label{eq:pi_meson}\\
 & \sigma|_{\theta=0} &  & =\bar{\psi}\psi &  & (J^{PG}=0^{++}),\label{eq:sigma_meson}\\
 & \eta|_{\theta=0} &  & =-i\bar{\psi}\gamma^{5}\psi &  & (J^{PG}=0^{--}).
 \label{eq:eta_meson}
\end{alignat}
Here, $J^{PG}$ show their isospin, parity, and $G$-parity quantum numbers.
When there is no room for confusion, we abbreviate the pions $\pi_{a}$ as $\pi$ for simplicity. 
For $m\ll g$, the masses of the pion and sigma meson scale as $(m/g)^{2/3}$, 
and they have the specific ratio $M_\sigma/M_\pi=\sqrt{3}$~\cite{Coleman:1976uz}. 
Thus, the decay process $\sigma \to \pi\pi$ is energetically forbidden, 
and the sigma meson is a stable particle, unlike the case of $(3+1)$d QCD. 
On the other hand, the eta meson has a heavier mass, $M_\eta\sim \mu=\sqrt{\frac{2}{\pi}}g$, 
compared with the pion and sigma meson, but it is stable due to the parity and $G$-parity quantum numbers. 
In the next subsection, we discuss how the $\theta$ term changes the physics discussed here.

\subsection{The particle spectrum at \texorpdfstring{$\theta \neq 0$}{nonzero theta}}
\label{subsec:spectrum_bosonization}

The $\theta$ term has an important effect on the low-energy physics. 
The structure of the internal global symmetry is strongly affected by the value of $\theta$, and it can be summarized as 
\begin{equation}
    \frac{\SU(2)_V}{\mathbb{Z}_2}\times\left\{
    \begin{array}{ll}
     (\mathbb{Z}_2)_G     & \quad (\theta=0), \\
     \mathrm{nothing}  &  \quad (\theta\not = 0,\pi), \\
     (\mathbb{Z}_2)_{\mathrm{chiral}+G} & \quad (\theta=\pi). 
    \end{array}
    \right.
    \label{eq:symmetry}
\end{equation}
The $\theta$ term violates the $G$-parity explicitly as it is odd under the charge conjugation, which makes the physics at $\theta=0$ special. 
We note that $\theta=\pi$ is another special point: 
Although the $G$-parity itself is violated, $(\mathbb{Z}_2)_{\mathrm{chiral}+G}\subset \SU(2)_L\times (\mathbb{Z}_2)_G$ is kept intact, 
where $\SU(2)_L$ is the chiral isospin transformation acting only on the left-moving fermions. 
In this section, we consider the case $0<\theta<\pi$ and postpone our discussion for $\theta=\pi$ to the next subsection. 

The Abelian bosonization gives the useful tool to analyze this system when $m\ll g$~\cite{Coleman:1976uz}, 
and the effective Lagrangian after integrating out the $U(1)$ gauge field becomes
\begin{equation}
    \mathcal{L}_{\mathrm{eff}}[\eta,\varphi]=\frac{1}{2}\left[(\partial\eta)^{2}-\mu^{2}\eta^{2}\right]+\frac{1}{4\pi}(\partial\varphi)^{2}+2Cm\rho N_{\rho}\left[\cos\left(\sqrt{2\pi}\eta-\frac{\theta}{2}\right)\cos\varphi\right],
    \label{eq:H_bosonization}
\end{equation}
where $\mu^{2}=2g^{2}/\pi$ and $C=e^{\gamma}/(2\pi)$, 
and the symbol $N_{\rho}[\cdot]$ denotes the normal ordering at the scale $\rho$~\cite{Coleman:1974bu}. 
Here $\eta$ is the non-compact scalar, and $\varphi$ is the $2\pi$-periodic scalar. 
In this description, the isospin $\SU(2)_V/\mathbb{Z}_2$ symmetry is not manifest in the Lagrangian, 
but it is realized as the symmetry acting on solitons at the quantum level. 
The $G$-parity is given by $\eta\to -\eta$, and we can explicitly see that it is a good symmetry only at $\theta=0$.
At $\theta=\pi$, the $G$-parity $\eta\to -\eta$ associated with $\varphi\to \varphi+\pi$ is the symmetry of this Lagrangian, 
and this is $(\mathbb{Z}_2)_{\mathrm{chiral}+G}$. 
We note that the parity $P$ is also violated for nonzero $\theta$, but $P'=P\times G$ is a good quantum number for any values of $\theta$. 

Since the $\eta$ field has the mass $\mu$ at the tree level, 
we can integrate it out to obtain the effective theory for the $2\pi$-periodic field $\varphi$. 
In the leading order, this gives the sine-Gordon theory, 
and we can analyze it by applying the optimized perturbation and the WKB analysis~\cite{Coleman:1976uz}. 
The lightest particle state turns out to be an iso-triplet pseudoscalar, 
which we identify as the pion, and its mass $M_\pi(\theta)$ is predicted to behave as
\begin{equation}
M_{\pi}(\theta)\sim\left|m\sqrt{\mu}\cos\frac{\theta}{2}\right|^{\frac{2}{3}}.
\label{eq:M_pi_theta}
\end{equation}
The next lightest state is an iso-singlet scalar, and we identify it as the $\sigma$ meson.
The WKB analysis gives a specific mass relation between the pion and sigma-meson masses, 
\begin{equation}
M_{\sigma}(\theta)=\sqrt{3}M_{\pi}(\theta).
\label{eq:sqrt3-Schwinger}
\end{equation}
Up to this point, the physics is the same between $\theta=0$ and nonzero $\theta$ 
except that the pion and sigma-meson masses have the nontrivial $\theta$ dependence. 

Let us move to the discussion on $\eta$. 
The computation of the self-energy for $\eta$ field is not so straightforward due to the potential infrared divergence~\cite{Coleman:1976uz}. 
We can still understand some heuristic behaviors of the eta meson from the symmetry and the effective Lagrangian. 
From the viewpoint of symmetry, the $\theta$ term violates both the parity and the $G$-parity, 
and thus $\eta$ is no longer distinguished from $\sigma$ or $\pi\pi$ scattering states by quantum numbers. 
This suggests the presence of the $\eta$-$\sigma$ mixing, $\eta\leftrightarrow \sigma$, 
and also the opening of the decay channel, $\eta \to \pi\pi$. 
Indeed, the interaction term of (\ref{eq:H_bosonization}) for $\eta$ gives
\begin{equation}
\cos\left(\sqrt{2\pi}\eta-\frac{\theta}{2}\right)\cos\varphi
=\left( \cos\frac{\theta}{2} + \sqrt{2\pi}\eta\sin\frac{\theta}{2} + \cdots \right)
\cos\varphi, 
\label{eq:eta_interaction}
\end{equation}
which contains the $\eta \cos\varphi$ vertex at $\theta \neq 0$.
Since the $\cos\varphi$ vertex creates the one-particle state of $\sigma$ and also the s-wave scattering states of $\pi\pi$, 
this gives the microscopic explanation of the $\eta$-$\sigma$ mixing and the $\eta\to \pi\pi$ decay.  

The scalar fields $\eta$ and $\varphi$ in \eqref{eq:H_bosonization} are related to the fermion bilinear operators as 
\begin{equation}
    \bar{\psi}_f\frac{1\pm \gamma_5}{2}\psi_f= - \frac{1}{2} C \rho N_\rho\left[\exp\left(\pm i\left(\sqrt{2\pi}\eta-\frac{\theta}{2}- (-1)^f \varphi\right) \right)\right],  
    \label{eq:fermionbilinear_bosonization}
\end{equation}
for each flavor $f=1,2$. 
This suggests that the meson operators, (\ref{eq:pi_meson}) - (\ref{eq:eta_meson}), 
should be rotated by the axial transformation to obtain the correct one-particle states, and we find
\begin{alignat}{3}
 & \pi_{a} &  & =-i\bar{\psi} \exp({i\frac{\theta}{2}\gamma^5})\gamma^{5}\tau_{a}\psi,
 \label{eq:pi_meson_theta}\\
 & \sigma &  & =\bar{\psi}\exp[{i\left(\frac{\theta}{2}+\omega(\theta)\right)\gamma^5}]\psi,
 \label{eq:sigma_meson_theta}\\
 & \eta &  & =-i\bar{\psi}\exp[{i\left(\frac{\theta}{2}+\omega(\theta)\right)\gamma^5}] \gamma^{5}\psi.
 \label{eq:eta_meson_theta}
\end{alignat}
For the sigma and eta mesons, the finite $m$ effect gives the mixing as we have seen in \eqref{eq:eta_interaction}, 
and the extra diagonalization is required, which is indicated by $\omega(\theta)$. 
The optimized perturbation suggests that $\omega(\theta)$ is of the order of $(m/g)^{4/3}\sin(\theta/2)\cos^{1/3}(\theta/2)$ 
at the leading order, so this extra rotation vanishes at $\theta=0$ and $\theta=\pi$. 
For the pion, there is no mixing counterpart due to the absence of the iso-triplet scalar particle, 
and thus such an extra rotation would not exist. 
We will numerically confirm this point in section~\ref{subsec:correlation_matrix}.

\subsection{Nearly conformal behaviors at \texorpdfstring{$\theta=\pi$}{theta=pi}}
\label{subsec:theory_at_critical_point}

Finally, let us consider the case of $\theta=\pi$.
When we set $m=0$ in \eqref{eq:H_bosonization}, 
we obtain the self-dual compact boson as the low-energy effective theory, which describes the $SU(2)_1$ WZW model. 
This is a conformal field theory that has the central charge $c=1$ and enjoys the $[\SU(2)_L\times \SU(2)_R]/\mathbb{Z}_2$ chiral symmetry. 
Turning on the small $m>0$ at $\theta=\pi$ gives the symmetry-breaking perturbation that preserves 
$[\SU(2)_V/\mathbb{Z}_2]\times (\mathbb{Z}_2)_{\mathrm{chiral}+G}$, and the possible lowest-dimensional scalar operator is given by $J_L J_R$. 
Here, $J_L$ and $J_R$ are holomorphic and anti-holomorphic parts of $SU(2)\times SU(2)$ chiral symmetry generators, 
and $J_L J_R$ is the scalar operator with the scaling dimension $2$. 
This $J_L J_R$ operator is non-integrably marginal, and thus it can be either marginally relevant or irrelevant 
depending on the sign of the coupling.\footnote{For the $2$d $\mathbb{C}P^1$ sigma model at $\theta=\pi$, 
the sign of the coupling is on the marginally irrelevant side and it flows back to the $SU(2)_1$ WZW fixed point logarithmically 
according to the Haldane conjecture~\cite{Haldane:1983ru, Affleck:1987ch}.} 
For the $2$-flavor Schwinger model at $\theta=\pi$, 
we can determine its sign by going beyond the leading-order analysis when integrating out $\eta$, 
and it turns out to be on the marginally relevant side~\cite{Coleman:1976uz}.
As a result, the system acquires the exponentially small mass gap $\sim e^{-\# g^{2}/m^{2}}g$, 
and the ground states are doubly degenerate due to the spontaneous breaking of $(\mathbb{Z}_2)_{\mathrm{chiral}+G}$~\cite{Dempsey:2023gib}.

In the numerical analysis, we put the system in a finite interval of size $L$. 
This introduces the typical energy scale of $O(\pi/L)$, 
and thus it is hard to detect a tiny mass gap unless we can take an exponentially large system size $L\gtrsim e^{\# g^{2}/m^{2}}/g $. 
We expect that the model at $\theta=\pi$ looks almost like a gapless system in our simulation setup, 
and thus the $SU(2)_1$ WZW model gives a good approximation.
Here, we summarize the analytic results of the one-point function with this approximation. 
The details of the calculation are summarized in appendix~\ref{sec:compact_boson_interval}.

Let us consider the open interval, $0\leq x\leq L$, and assume that we put the Dirichlet boundary condition in the bosonized description, 
\begin{equation}
    \eta(x=0,L)=0, \quad \varphi(x=0,L)=0. 
\end{equation}
This can be translated to the boundary condition for fermions by using \eqref{eq:fermionbilinear_bosonization}. 
We then find that 
\begin{align}
    &\Braket{\sigma(x)} = \langle \overline{\psi}i \gamma_5 \psi (x)\rangle 
    = - \sqrt{ \frac{e^\gamma \mu}{\pi L} } \frac{1}{\sqrt{\sin (\pi x/L)}} , 
    \label{eq:cft_sigma} 
    \\
    &\Braket{\pi(x)} = \langle \bar{\psi}\tau_3\psi(x)\rangle =0. 
    \label{eq:cft_pi}
\end{align}
The result $\Braket{\pi(x)}=0$ follows immediately from the fact that the boundary condition is the isospin symmetric. 

We can easily generalize the result to the isospin-violating Dirichlet boundary condition by introducing the twist angle $\Delta$. 
For example, if we take 
\begin{equation}
    \varphi(x=0, L)=-\Delta,
    \label{eq:twist_bc_even}
\end{equation}
the analytic forms of the sigma and pion one-point functions are given by 
\begin{align}
    &\Braket{\sigma(x)}
    = - \sqrt{ \frac{e^\gamma \mu}{\pi L} } \frac{\cos \Delta}{\sqrt{\sin (\pi x/L)}} , 
    \label{eq:cft_sigma_even} 
    \\
    &\Braket{\pi(x)} = \sqrt{ \frac{e^\gamma \mu}{\pi L} } \frac{\sin \Delta}{\sqrt{\sin (\pi x/L)}}. 
    \label{eq:cft_pi_even}
\end{align}
As another isospin-violating Dirichlet boundary condition, we may take 
\begin{equation}
    \varphi(x=0)=-\varphi(x=L)=-\Delta,
    \label{eq:twist_bc_odd}
\end{equation}
and we here assume $|\Delta|<\pi/2$. Then, we obtain 
\begin{align}
    &\Braket{\sigma(x)}
    = - \sqrt{ \frac{e^\gamma \mu}{\pi L} } \frac{\cos [\Delta(1-2x/L)]}{\sqrt{\sin (\pi x/L)}} , 
    \label{eq:cft_sigma_odd} 
    \\
    &\Braket{\pi(x)} = \sqrt{ \frac{e^\gamma \mu}{\pi L} } \frac{\sin [\Delta(1-2x/L)]}{\sqrt{\sin (\pi x/L)}}. 
    \label{eq:cft_pi_odd}
\end{align}
The result has a nontrivial dependence on the angle $\Delta$.

\section{Basic calculation strategy and review at \texorpdfstring{$\theta=0$}{theta=0}}
\label{sec:strategy}

Here, we address the basics of calculation and explain our strategy to obtain the mass spectra 
in the previous work at $\theta=0$~\cite{Itou:2023img}.
First, we define the Hamiltonian of the $2$-flavor Schwinger model and its lattice regularization 
and explain how to generate the ground state in the DMRG method. 
After that, we quickly review three computational schemes for the meson spectra, 
which were demonstrated in the $2$-flavor Schwinger model at $\theta=0$ in our previous work, 
and summarize each pros and cons found in the actual calculations.

\subsection{Lattice Hamiltonian simulation of the 2-flavor Schwinger model with the open boundary condition}
\label{subsec:lattice_Hamiltonian}

We use the lattice Hamiltonian formalism to perform the numerical computation of the $2$-flavor Schwinger model~\eqref{eq:Lagrangian}. 
As we use the same formalism with our previous paper~\cite{Itou:2023img}, we keep our explanation brief when it has an overlap. 

We put the system on the open interval $0\leq x\leq L$, and perform the canonical quantization in the temporal gauge. 
In this setup, we can solve the Gauss law constraint explicitly, 
and the Hamiltonian can be completely described by fermions with non-local interactions. 
We take the lattice discretization of this fermionic system using the staggered fermion~\cite{Kogut:1974ag,Susskind:1976jm} 
and obtain the quantum system with the finite-dimensional Hilbert space. 
Let $\chi_{f,n}$ be the staggered fermion on the lattice sites $n=0,1,\cdots,N-1$ with the lattice spacing $a$ (i.e. $(N-1)a=L$), 
and then it is translated to the continuum Dirac fermion $\psi_f(x)$ at $x=na$ as 
\begin{equation}
    \psi_f(x)\leftrightarrow 
    \frac{1}{\sqrt{2a}}\begin{pmatrix}
        \chi_{f,2[n/2]}\\
        \chi_{f,2[n/2]+1}
    \end{pmatrix}. 
    \label{eq:fermion_dictionary}
\end{equation}
We use this dictionary to obtain the lattice expression for various fermion bilinear operators, 
and we give explicit forms for important ones in appendix~\ref{sec:operators}. 
After these manipulations, the lattice Hamiltonian of $N_f$-flavor Schwinger model consists of the gauge part $H_{J}$, 
the fermion kinetic term $H_{w}$, and the mass term $H_{m}$, i.e. $H=H_J+H_w+H_m$, and each term is given by 
\begin{align}
    H_{J}&=J\sum_{n=0}^{N-2}\left[\sum_{f=1}^{N_{f}}\sum_{k=0}^{n}\chi_{f,k}^{\dagger}\chi_{f,k}+\frac{N_{f}}{2}\left(\frac{(-1)^{n}-1}{2}-n\right)+\frac{\theta}{2\pi}\right]^{2},
    \label{eq:H_J} \\
    H_{w}&=-iw\sum_{f=1}^{N_{f}}\sum_{n=0}^{N-2}\left(\chi_{f,n}^{\dagger}\chi_{f,n+1}-\chi_{f,n+1}^{\dagger}\chi_{f,n}\right), 
    \label{eq:H_w} \\
    H_{m}&=m_{\mathrm{lat}}\sum_{f=1}^{N_{f}}\sum_{n=0}^{N-1}(-1)^{n}\chi_{f,n}^{\dagger}\chi_{f,n},
    \label{eq:H_m}
\end{align}
where $J=g^{2}a/2$ and $w=1/2a$. 
We relate the lattice fermion mass $m_{\mathrm{lat}}$ and the mass $m$ of the continuum theory as 
\begin{equation}
m_{\mathrm{lat}}:=m-\frac{N_{f}g^{2}a}{8},
\label{eq:latticemass}
\end{equation}
so that we maintain the $\mathbb{Z}_2$ discrete chiral symmetry in the chiral limit 
and prevent the additive renormalization~\cite{Dempsey:2022nys}.

\subsection{Generation of the ground state}

When performing the actual computation, we further map this fermionic system to the spin system via the Jordan-Winger transformation, 
and its details are explained in section~3 and appendix~A of~\cite{Itou:2023img} with the same notation. 
After this mapping, we apply the density-matrix renormalization group (DMRG)~\cite{White:1992zz,White:1993zza,Schollw_ck_2005,Schollw_ck_2011}. 

The DMRG is a variational algorithm using the matrix product state (MPS) as an ansatz of the wave function.
It can find the ground state, which minimizes the cost function, namely the expectation value of the Hamiltonian.
There is a cutoff parameter $\varepsilon$ controlling the truncation error of the singular-value decomposition (SVD), 
which determines the bond dimension $D$ of MPS.
Smaller $\varepsilon$ yields a better approximation, whereas it requires a larger bond dimension and a higher computational cost.
We choose the N\'{e}el state as an initial state of the DMRG, which corresponds to a zero-particle state in terms of the Schwinger model.
We also impose the conservation of the total electric charge $Q$ during the DMRG, 
so that the resulting MPS satisfies $Q=0$ as required by the Gauss law. 

\begin{figure}[htb]
\centering
\includegraphics[scale=0.5]{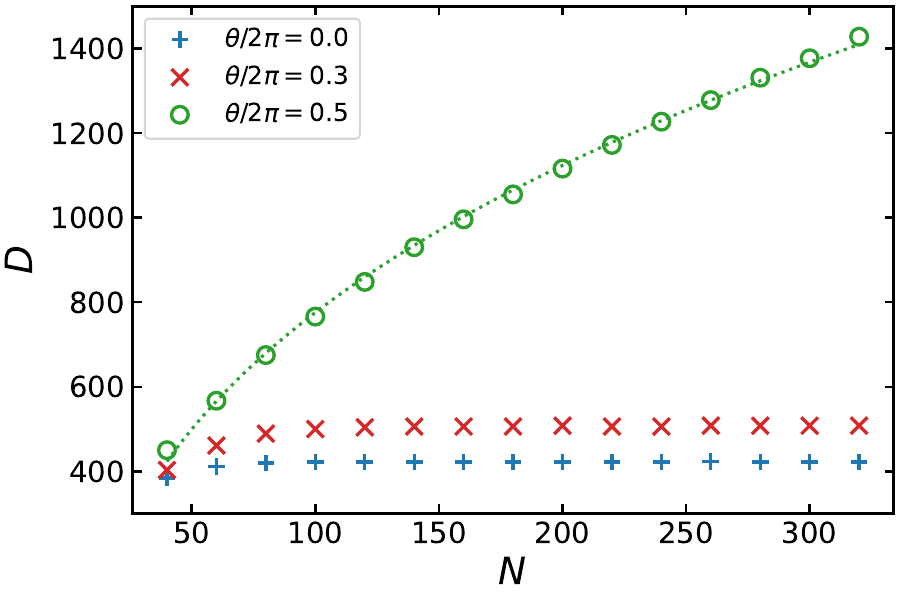}
\caption{\label{fig:bond_dim_a025}
The bond dimension $D$ of the MPS is plotted against the system size $N$
for $\theta/2\pi=0.0$, $0.3$, and $0.5$.
The lattice spacing is set to $a=0.25$.
The data for $\theta/2\pi=0.5$ are fitted by $c_1 N^{1/3} + c_2$, 
where the dotted curve denotes the fitting result.}
\end{figure}

Let us comment on the computational cost of DMRG when applying it to the $2$-flavor Schwinger model at $\theta\neq 0$.
The potential difficulty comes from the increase of the bond dimension, 
because the gap of the system, namely the pion mass, decreases as $\theta$ approaches $\pi$.
At $\theta=\pi$, the system becomes almost gapless, 
and the entanglement entropy is no longer constant but increases with the lattice size $N$ 
as $S_{\mathrm{EE}} \sim (c/3)\log N$ with the central charge $c=1$ in the continuum limit.
Correspondingly, the required bond dimension $D$ is expected to increase as $D \sim N^{c/3}$.
Figure~\ref{fig:bond_dim_a025} depicts the example plot of the bond dimension $D$ of the MPS for the ground state as a function of $N$.
Here, we set the theoretical parameter to $m/g=0.10$ and the lattice spacing to $a=0.25$, 
which are the same as the ones in the actual calculation in this study.
While $D$ is saturated to a constant value in the gapped cases of $0\leq\theta<\pi$, it increases with $N$ at $\theta=\pi$.
By fitting $D$ at $\theta=\pi$ with $c_1 N^{1/3} + c_2$, we obtain $c_1=288(3)$ and $c_2=-563(16)$.
Note that a finite spacing effect can contribute to a discrepancy between the data and the fitting result.
In this study, we take $N\approx 300$, then the necessary bond dimension is $D \lesssim 1400$. 
It is a doable calculation using a PC cluster or a single-node supercomputer.

\subsection{Review of computational schemes for the meson spectra at \texorpdfstring{$\theta=0$}{theta=0}}
\label{sec:schemes}

In our previous work~\cite{Itou:2023img}, we developed three distinct methods to compute the mass spectrum in the Hamiltonian formalism.
We demonstrated the methods in the $2$-flavor Schwinger model at $\theta=0$ using DMRG 
and confirmed that the results are consistent with each other.
Let us briefly summarize the features of the three methods.

\subsubsection{Correlation-function scheme}

The first one, the correlation-function scheme, is to employ the technique in the Lagrangian formalism,
where we obtain the meson mass from the spatial two-point correlation function for desired composite states.
In Euclidean path integral, 
the meson mass can be calculated from the imaginary-time two-point correlation function, 
and it is equivalent to the spatial two-point correlation function thanks to the Lorentz invariance: 
When an operator $O(r)$ creates the one-particle state with the mass $M$, 
the asymptotic behavior of the spatial correlator takes the Yukawa-type form,
\beq
\langle \mathcal{O}(r) \mathcal{O}(0)\rangle -\langle O(r)\rangle \langle O(0)\rangle \sim \frac{1}{r^\alpha} \exp(-Mr).
\label{eq:cf_asymptotic}
\eeq
Strictly speaking, the imaginary-time correlator and the spatial correlator are no longer equivalent 
in the lattice regularized Hamiltonian formalism, and the equivalence is recovered after taking the continuum limit. 

While this method can apply to various models in any dimension, there is a subtlety when using the tensor network. 
In eq.~\eqref{eq:cf_asymptotic}, there is the power-like behavior $1/r^\alpha$ in front of the exponential factor, 
where $\alpha=(d-1)/2$ for the free scalar with the spacetime dimension $d$. 
However, any tensor-network-based algorithms are a kind of finite-rank approximation of the transfer matrix, 
and reproducing such a power-law factor requires a large bond dimension.
Thus, we need to increase the bond dimension until the correct asymptotic behavior is observed.

Let us point out another issue when we take the open boundary condition. 
The functional form~\eqref{eq:cf_asymptotic} assumes the translational invariance, but the presence of boundaries violates it. 
Therefore, we need to take a sufficiently large system size 
so that the insertion of local operators is very far from the boundaries to make their effect negligible. 
When we consider the situation of the tiny mass gap, this is difficult to achieve, 
and we have to take careful extrapolations to find reliable results. 
In this paper, we find it difficult to obtain the mass spectra at larger values of $\theta$, 
so we put the analysis of the correlation-function scheme in appendix~\ref{sec:cf_scheme} instead of the main text. 

On the other hand, the correlation function still provides the most systematic way to analyze the mixing of states.
By considering the set of local operators $O_i(r)$, we can compute the matrix of correlation function $\langle O_i(r) O_j(0) \rangle$, 
and we may extract the information of excited states, state-mixing, etc. by its diagonalization. 
This is the unique feature of the correlation-function scheme, which is absent in the other two methods.

\subsubsection{One-point-function scheme}

The second method, the one-point-function scheme, 
utilizes the boundary or defect as a source for excitations from the thermodynamic ground state. 
Let us again assume that the Lorentz symmetry is approximately recovered, 
then we may regard the spatial direction as the imaginary-time direction. 
Then, the boundary gives some state, $|\mathrm{Bdry}\rangle$, and the one-point function can be interpreted as 
\begin{equation}
    \langle O(x)\rangle = \langle \mathrm{Bdry}| e^{-H x} O |\Omega\rangle, 
\end{equation}
where $|\Omega\rangle$ is the ground state and we normalize the ground-state energy to be $0$. 
As $|\mathrm{Bdry}\rangle$ is translational invariant, it is the zero mode of the momentum operator, 
and the asymptotic behavior of the one-point function for the gapped system should behave as 
\beq
\langle \mathcal{O}(x) \rangle \sim e^{-Mx},
\label{eq:one_point_function}
\eeq
where $M$ is the mass of the lightest meson with the quantum number of $O$. 
This is not the Yukawa-type but the simple exponential decay unlike eq.~\eqref{eq:cf_asymptotic}, 
and one may understand it with the analogy of the wall source method in lattice QCD.

We often need to choose the symmetry-violating boundary condition when we measure the meson mass with the nontrivial quantum number. 
Otherwise, the one-point function vanishes due to the symmetry, and we cannot measure the mass of the target particle. 
For example, in the previous work~\cite{Itou:2023img}, we used the edge mode of the SPT state as a source of the triplet mesons 
by setting $\theta=2\pi$, because the naive open boundary condition at $\theta=0$ is isospin singlet and cannot be used to measure the pion mass.

Let us comment that the simple exponential form in \eqref{eq:one_point_function} has nice compatibility with the tensor network algorithms. 
We find that the one-point function is insensitive to the bond dimension, 
which allows us to set a larger $\varepsilon$ than that in the correlation-function scheme. 
However, one cannot access the information of the off-diagonal correlators unlike the correlation-function scheme, 
so this method is specialized to the lightest particle spectrum in a given quantum number.

\subsubsection{Dispersion-relation scheme}

The third method, the dispersion-relation scheme, is the distinctive strategy of Hamiltonian formalism.
We generated the low-energy excited states as well as the ground state.
Suppose we already have the lower energy eigenstates $\ket{\Psi_{\ell'}}$ for the level $\ell'=0,1,\cdots,\ell-1$. 
Then the $\ell$-th excited state $\ket{\Psi_{\ell}}$ can be obtained by DMRG with the Hamiltonian modified as 
\begin{equation}
H_{\ell}=H+W\sum_{\ell^{\prime}=0}^{\ell-1}\ket{\Psi_{\ell^{\prime}}}\bra{\Psi_{\ell^{\prime}}},
\label{eq:H_excited_states}
\end{equation}
where the parameter $W>0$ is a weight to impose the orthogonality~\cite{Wall_2012,Banuls:2013jaa}.

The type of the meson was identified by the quantum numbers, such as the isospin and $G$-parity.
After identifying the state of each meson by the quantum numbers, the mass is obtained by the dispersion relation, $E=\sqrt{M^2+K^2}$, 
for each meson by measuring the energy ($E$) and momentum square ($K^2$).
This is a basic idea of this scheme. 

As we work on the open boundary, 
the translational invariance is explicitly broken and the momentum operator does not provide a good quantum number. 
As an ad hoc substitute, we identify the momentum square of the $\ell$-th state by subtracting the ground-state contribution, 
$\Delta K^2_\ell:=\langle \Psi_\ell| K^2 |\Psi_\ell\rangle - \langle \Psi_0| K^2 |\Psi_0\rangle$, and it works well empirically.

This method is a heuristic approach and allows the detection of states other than single mesons.
This paper does not address such states, but they are considered to be scattering states~\cite{Harada:1993va}.
On the other hand, the efficiency of this method depends on the way of obtaining the excited states of interest.
If we are interested in the high-level state, 
the computational cost of finding the target state from the others can be a bottleneck.

\section{Caluclation strategy for \texorpdfstring{$\theta \neq 0$}{nonzero theta} regime}
\label{sec:strategy_for_theta}

At $\theta \neq 0$, the ground state can be obtained by DMRG, 
but the expected physics are different from the one at $\theta = 0$ as discussed 
in sections~\ref{subsec:spectrum_bosonization} and~\ref{subsec:theory_at_critical_point}.
For instance, there is an operator mixing between the scalar and pseudo-scalar in the state, and some quantum numbers are no longer exact. 
We have to resolve the mixing between them to obtain the mass eigenstate. 
Furthermore, a nearly conformal theory emerges at $\theta=\pi$ where the functional form of the one-point function must be changed 
as shown in section~\ref{subsec:theory_at_critical_point}.
Therefore, some methodological improvements are needed.

From these situations, we take two types of calculation and introduce some technical improvements at $\theta\neq 0$ 
to obtain the promising result of the mass spectrum.
The first method is an improved one-point-function scheme combining the correlation-function and one-point-function schemes at $\theta=0$. 
In this scheme, we introduce an alternative boundary condition to control the boundary effects.
The second one is the dispersion-relation scheme, which is the same as the one at $\theta=0$. 
However, the parity and $G$-parity are no longer the quantum numbers of the mesons for $\theta \neq 0$,
since the $\theta$ term explicitly breaks these symmetries.
Moreover, the mass spectra are getting lighter in the large $\theta$ regime, 
so that the computational cost to obtain the excited states increases. 
Therefore, we also improved the technique to generate excited states in the DMRG calculation.

\subsection{Improved one-point-function scheme}

The basic idea of this scheme is the following:
First of all, we measure the two-point correlation function for the $2 \times 2$ matrix of the scalar and pseudo-scalar operators 
and find the mixing angle between them by diagonalizing the matrix. 
Then, we evaluate the one-point function of the bulk operator with the mass eigenstates. 
Technically, to obtain the precise value of the mass, we introduce some modified boundary conditions by twisting fermion masses.

\subsubsection{Determination of the mixing matrix}
\label{sec:def-mixing-matrix}

To resolve the operator mixing, we measure the two-point correlation function for the $2 \times 2$ matrix,
\begin{equation}
\boldsymbol{C}_{\pm}(x,y)=
\begin{pmatrix}
\Braket{S_{\pm}(x)S_{\pm}(y)}_{c} & \Braket{S_{\pm}(x)PS_{\pm}(y)}_{c}\\
\Braket{PS_{\pm}(x)S_{\pm}(y)}_{c} & \Braket{PS_{\pm}(x)PS_{\pm}(y)}_{c}
\end{pmatrix}, 
\label{eq:C_mat}
\end{equation}
for the iso-triplet and iso-singlet sectors, where the subscript $c$ means the connected correlator, 
$\Braket{\mathcal{O}_{1}\mathcal{O}_{2}}_{c}
=\Braket{\mathcal{O}_{1}\mathcal{O}_{2}}-\Braket{\mathcal{O}_{1}}\Braket{\mathcal{O}_{2}}$.
Here, we introduce the lattice version of the iso-triplet and iso-singlet fermion bilinear operators: 
The $\tau_3$-components of the iso-triplet one are denoted as 
\begin{equation}
    S_{-}(x) \leftrightarrow \overline{\psi}\tau_3\psi(x),\quad 
    PS_{-}(x) \leftrightarrow -i \overline{\psi}\gamma^5\tau_3\psi(x), 
\end{equation}
with $x=na$, and the iso-singlet ones are denoted as 
\begin{equation}
    S_{+}(x) \leftrightarrow \overline{\psi}\psi(x),\quad 
    PS_{+}(x) \leftrightarrow -i \overline{\psi}\gamma^5\psi(x). 
\end{equation}
The actual expression for the lattice fermion is obtained by using the dictionary~\eqref{eq:fermion_dictionary} 
and taking the three-lattice-point average to smear the oscillation coming from the use of staggered fermions. 

Since $S_{\pm}(x)$ and $PS_{\pm}(x)$ are Hermitian operators, 
$\boldsymbol{C}_{\pm}(x,y)$ is a real symmetric matrix, which can be diagonalized by an orthogonal matrix, 
\begin{equation}
    R(\delta) = 
    \begin{pmatrix}
    \cos\delta & -\sin\delta\\
    \sin\delta & \cos\delta
    \end{pmatrix}. \label{eq:def_R}
\end{equation}
Here the argument $\delta$ corresponds to the mixing angle, 
and the eigenvalues of $\boldsymbol{C}_{\pm}(x,y)$ are the connected correlation functions of the meson operators.
Thus, we define the operators $\sigma(x)$ and $\eta(x)$ by 
\begin{equation}
    \boldsymbol{C}_{+}(x,y)=
    R(\delta_+)^{\mathrm{T}}
    \begin{pmatrix}
    \Braket{\sigma(x)\sigma(y)}_{c} & 0\\
    0 & \Braket{\eta(x)\eta(y)}_{c}
    \end{pmatrix}
    R(\delta_+),
\end{equation}
and the operator $\pi(x)$ by
\begin{equation}
    \boldsymbol{C}_{-}(x,y)=
    R(\delta_-)^{\mathrm{T}}
    \begin{pmatrix}
    \hphantom{\pi(x)}*\hphantom{\pi(y)} & 0\\
    0 & \Braket{\pi(x)\pi(y)}_{c}
    \end{pmatrix}
    R(\delta_-).
\end{equation}
The explicit forms for $\pi(x)$, $\sigma(x)$ and $\eta(x)$ are given by 
\begin{equation}
\begin{pmatrix}*\\
\pi(x)
\end{pmatrix}=R(\delta_-)\begin{pmatrix}S_{-}(x)\\
PS_{-}(x)
\end{pmatrix},
\qquad 
\begin{pmatrix}\sigma(x)\\
\eta(x)
\end{pmatrix}=R(\delta_+)\begin{pmatrix}S_{+}(x)\\
PS_{+}(x)
\end{pmatrix}.
\label{eq:meson_op}
\end{equation}
Note that $\delta_{\pm}$ could depend on the distance $r=|x-y|$ 
at which $\boldsymbol{C}_{\pm}(x,y)$ is measured due to the boundary effect.
However, we find from actual simulation results that $\delta_{\pm}$ is almost independent of $r$ in the bulk region, 
thus such a boundary effect on $\delta_{\pm}$ is sufficiently small.
We also confirm that $\delta_{\pm}$ is not sensitive to the cutoff parameter $\varepsilon$.

Theoretically, we can estimate the mixing angle as follows.
The finite-$m$ interaction term induces the presence of the $\eta$-$\sigma$ mixing. 
Using the discussion around \eqref{eq:eta_interaction}, 
we can give the analytic form for $\delta_+$ from the bosonized effective Lagrangian. 
The $\eta$-$\sigma$ mixing is characterized by the mass matrix, 
\begin{equation}
\mathcal{M}(\theta) = \mu^2 \begin{pmatrix}
1 & A\sin(\theta/2)|\cos(\theta/2)|^{1/3} \\
A\sin(\theta/2)|\cos(\theta/2)|^{1/3} & B|\cos(\theta/2)|^{4/3}
\end{pmatrix}, \label{eq:M-theta}
\end{equation}
where $A$ and $B$ are $O((m/g)^{4/3})$ quantities.
The real-symmetric matrix $\mathcal{M}(\theta)$ can be diagonalized 
by the orthogonal matrix $R(\omega(\theta))$ with the argument $\omega(\theta)$, 
which corresponds to the extra rotation in eqs.~\eqref{eq:sigma_meson_theta} and~\eqref{eq:eta_meson_theta}.
Then, $\delta_{-} \simeq \theta/2$ and $\delta_{+} \simeq \theta/2 + \omega(\theta)$ are expected by the bosonization analysis.

\subsubsection{Modifying the boundary and twisting fermion mass}
\label{subsec:twisted_mass}

Before measuring the bulk one-point function for each meson given in eq.~\eqref{eq:meson_op}, 
let us introduce an alternative boundary condition instead of the naive open boundary.
The key idea of the one-point-function scheme is the usage of the edge mode that appears at the boundary as the source of the meson.
An alternative boundary condition makes it possible to change the source operator more flexibly.

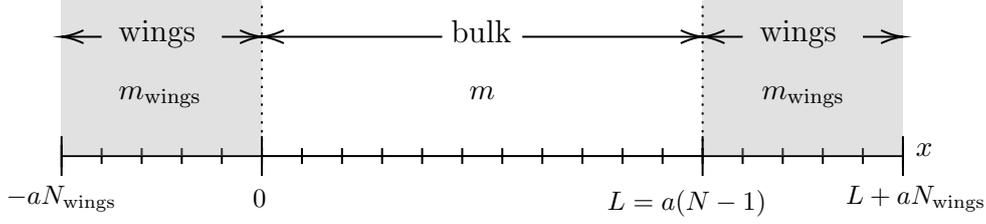
\begin{figure}[htb]
\centering
\tikzset{every picture/.style={line width=0.75pt}} 

\begin{tikzpicture}[x=0.75pt,y=0.75pt,yscale=-1,xscale=1]

\draw [line width=0.75]    (120,190) -- (540,190) (140,186) -- (140,194)(160,186) -- (160,194)(180,186) -- (180,194)(200,186) -- (200,194)(220,186) -- (220,194)(240,186) -- (240,194)(260,186) -- (260,194)(280,186) -- (280,194)(300,186) -- (300,194)(320,186) -- (320,194)(340,186) -- (340,194)(360,186) -- (360,194)(380,186) -- (380,194)(400,186) -- (400,194)(420,186) -- (420,194)(440,186) -- (440,194)(460,186) -- (460,194)(480,186) -- (480,194)(500,186) -- (500,194)(520,186) -- (520,194) ;
\draw [shift={(540,190)}, rotate = 180] [color={rgb, 255:red, 0; green, 0; blue, 0 }  ][line width=0.75]    (0,5.59) -- (0,-5.59)   ;
\draw [shift={(120,190)}, rotate = 180] [color={rgb, 255:red, 0; green, 0; blue, 0 }  ][line width=0.75]    (0,5.59) -- (0,-5.59)   ;
\draw  [draw opacity=0][fill={rgb, 255:red, 155; green, 155; blue, 155 }  ,fill opacity=0.3 ] (120,110) -- (220,110) -- (220,190) -- (120,190) -- cycle ;
\draw  [draw opacity=0][fill={rgb, 255:red, 155; green, 155; blue, 155 }  ,fill opacity=0.3 ] (440,110) -- (540,110) -- (540,190) -- (440,190) -- cycle ;
\draw  [dash pattern={on 0.84pt off 2.51pt}]  (220,110) -- (220,190) ;
\draw  [dash pattern={on 0.84pt off 2.51pt}]  (440,110) -- (440,190) ;
\draw    (440,180) -- (440,200) ;
\draw    (220,180) -- (220,200) ;
\draw    (120,180) -- (120,200) ;
\draw    (540,180) -- (540,200) ;
\draw    (200,130) -- (218,130) ;
\draw [shift={(220,130)}, rotate = 180] [color={rgb, 255:red, 0; green, 0; blue, 0 }  ][line width=0.75]    (10.93,-3.29) .. controls (6.95,-1.4) and (3.31,-0.3) .. (0,0) .. controls (3.31,0.3) and (6.95,1.4) .. (10.93,3.29)   ;
\draw    (520,130) -- (538,130) ;
\draw [shift={(540,130)}, rotate = 180] [color={rgb, 255:red, 0; green, 0; blue, 0 }  ][line width=0.75]    (10.93,-3.29) .. controls (6.95,-1.4) and (3.31,-0.3) .. (0,0) .. controls (3.31,0.3) and (6.95,1.4) .. (10.93,3.29)   ;
\draw    (140,130) -- (122,130) ;
\draw [shift={(120,130)}, rotate = 360] [color={rgb, 255:red, 0; green, 0; blue, 0 }  ][line width=0.75]    (10.93,-3.29) .. controls (6.95,-1.4) and (3.31,-0.3) .. (0,0) .. controls (3.31,0.3) and (6.95,1.4) .. (10.93,3.29)   ;
\draw    (460,130) -- (442,130) ;
\draw [shift={(440,130)}, rotate = 360] [color={rgb, 255:red, 0; green, 0; blue, 0 }  ][line width=0.75]    (10.93,-3.29) .. controls (6.95,-1.4) and (3.31,-0.3) .. (0,0) .. controls (3.31,0.3) and (6.95,1.4) .. (10.93,3.29)   ;
\draw    (350,130) -- (438,130) ;
\draw [shift={(440,130)}, rotate = 180] [color={rgb, 255:red, 0; green, 0; blue, 0 }  ][line width=0.75]    (10.93,-3.29) .. controls (6.95,-1.4) and (3.31,-0.3) .. (0,0) .. controls (3.31,0.3) and (6.95,1.4) .. (10.93,3.29)   ;
\draw    (310,130) -- (222,130) ;
\draw [shift={(220,130)}, rotate = 360] [color={rgb, 255:red, 0; green, 0; blue, 0 }  ][line width=0.75]    (10.93,-3.29) .. controls (6.95,-1.4) and (3.31,-0.3) .. (0,0) .. controls (3.31,0.3) and (6.95,1.4) .. (10.93,3.29)   ;

\draw (214,205.4) node [anchor=north west][inner sep=0.75pt]  [font=\small]  {$0$};
\draw (391,205.4) node [anchor=north west][inner sep=0.75pt]  [font=\small]  {$L=a( N-1)$};
\draw (91,203.4) node [anchor=north west][inner sep=0.75pt]  [font=\small]  {$-aN_{\mathrm{wings}}$};
\draw (510,203.4) node [anchor=north west][inner sep=0.75pt]  [font=\small]  {$L+aN_{\mathrm{wings}}$};
\draw (322,153.4) node [anchor=north west][inner sep=0.75pt]    {$m$};
\draw (147,153.4) node [anchor=north west][inner sep=0.75pt]    {$m_{\mathrm{wings}}$};
\draw (313,120) node [anchor=north west][inner sep=0.75pt]  [font=\large] [align=left] {bulk};
\draw (147,120) node [anchor=north west][inner sep=0.75pt]  [font=\large] [align=left] {wings};
\draw (467,120) node [anchor=north west][inner sep=0.75pt]  [font=\large] [align=left] {wings};
\draw (545,182.4) node [anchor=north west][inner sep=0.75pt]    {$x$};
\draw (468,153.4) node [anchor=north west][inner sep=0.75pt]    {$m_{\mathrm{wings}}$};

\end{tikzpicture}
\caption{The schematic picture of the lattice setup with the wings regime.
We assign the mass parameter $m$ for $0 \leq x \leq L$ and $m_\mathrm{wings}$ for $x<0$ and $L<x$.}
\label{fig:wings-image}
\end{figure}

Now, we consider supplemental lattice sites to both ends of the $1$d lattice.
We refer to the attached region as the wings regime~\footnote{
It comes from the term {\it stage wings} in theater, namely the hidden parts at each side of a stage.}.
Then, we assign the heavier fermion mass in the wings regime than that in the bulk, 
namely $m \ll m_{\rm wings}$ as shown in figure~\ref{fig:wings-image}.
This setup corresponds to the Dirichlet boundary condition for the Dirac fermion 
and is easily implemented by changing the mass $m$ locally in the Hamiltonian~\eqref{eq:H_m}.
The boundary between the bulk and the wings becomes the source of mesons.
This is a basic idea of the alternative boundary condition.

In the case of the iso-singlet sector, namely the sigma and eta mesons, we simply set the large constant,
$m_{\rm wings}=m_0$, as the mass in the wings regime.
Similarly to the edge mode under the naive open boundary condition that results in an iso-singlet source operator at $\theta=0$, 
the bulk one-point function of the iso-singlet operators can be obtained with the simple setup even at $\theta \neq 0$.

As for the iso-triplet sector, we apply a flavor-asymmetric twist to the fermion mass in the wings regime~\footnote{
We utilized the property of the SPT phase as the source of the triplet meson by shifting $\theta \rightarrow \theta + 2\pi$ in our previous work. 
Similar analyses are shown in appendix~\ref{sec:1pt_func_original}.}.
We define the twisted mass by acting chiral rotation with the opposite phase $\pm\Delta$ for the flavor $f=1,2$.
\begin{figure}[htb]
\centering
\tikzset{every picture/.style={line width=0.75pt}} 

\begin{tikzpicture}[x=0.75pt,y=0.75pt,yscale=-1,xscale=1]

\draw [line width=0.75]    (120,190) -- (540,190) (140,186) -- (140,194)(160,186) -- (160,194)(180,186) -- (180,194)(200,186) -- (200,194)(220,186) -- (220,194)(240,186) -- (240,194)(260,186) -- (260,194)(280,186) -- (280,194)(300,186) -- (300,194)(320,186) -- (320,194)(340,186) -- (340,194)(360,186) -- (360,194)(380,186) -- (380,194)(400,186) -- (400,194)(420,186) -- (420,194)(440,186) -- (440,194)(460,186) -- (460,194)(480,186) -- (480,194)(500,186) -- (500,194)(520,186) -- (520,194) ;
\draw [shift={(540,190)}, rotate = 180] [color={rgb, 255:red, 0; green, 0; blue, 0 }  ][line width=0.75]    (0,5.59) -- (0,-5.59)   ;
\draw [shift={(120,190)}, rotate = 180] [color={rgb, 255:red, 0; green, 0; blue, 0 }  ][line width=0.75]    (0,5.59) -- (0,-5.59)   ;
\draw  [draw opacity=0][fill={rgb, 255:red, 155; green, 155; blue, 155 }  ,fill opacity=0.3 ] (120,110) -- (220,110) -- (220,190) -- (120,190) -- cycle ;
\draw  [draw opacity=0][fill={rgb, 255:red, 155; green, 155; blue, 155 }  ,fill opacity=0.3 ] (440,110) -- (540,110) -- (540,190) -- (440,190) -- cycle ;
\draw  [dash pattern={on 0.84pt off 2.51pt}]  (220,110) -- (220,190) ;
\draw  [dash pattern={on 0.84pt off 2.51pt}]  (440,110) -- (440,190) ;
\draw    (440,180) -- (440,200) ;
\draw    (220,180) -- (220,200) ;
\draw    (120,180) -- (120,200) ;
\draw    (540,180) -- (540,200) ;
\draw    (200,130) -- (218,130) ;
\draw [shift={(220,130)}, rotate = 180] [color={rgb, 255:red, 0; green, 0; blue, 0 }  ][line width=0.75]    (10.93,-3.29) .. controls (6.95,-1.4) and (3.31,-0.3) .. (0,0) .. controls (3.31,0.3) and (6.95,1.4) .. (10.93,3.29)   ;
\draw    (520,130) -- (538,130) ;
\draw [shift={(540,130)}, rotate = 180] [color={rgb, 255:red, 0; green, 0; blue, 0 }  ][line width=0.75]    (10.93,-3.29) .. controls (6.95,-1.4) and (3.31,-0.3) .. (0,0) .. controls (3.31,0.3) and (6.95,1.4) .. (10.93,3.29)   ;
\draw    (140,130) -- (122,130) ;
\draw [shift={(120,130)}, rotate = 360] [color={rgb, 255:red, 0; green, 0; blue, 0 }  ][line width=0.75]    (10.93,-3.29) .. controls (6.95,-1.4) and (3.31,-0.3) .. (0,0) .. controls (3.31,0.3) and (6.95,1.4) .. (10.93,3.29)   ;
\draw    (460,130) -- (442,130) ;
\draw [shift={(440,130)}, rotate = 360] [color={rgb, 255:red, 0; green, 0; blue, 0 }  ][line width=0.75]    (10.93,-3.29) .. controls (6.95,-1.4) and (3.31,-0.3) .. (0,0) .. controls (3.31,0.3) and (6.95,1.4) .. (10.93,3.29)   ;
\draw    (350,130) -- (438,130) ;
\draw [shift={(440,130)}, rotate = 180] [color={rgb, 255:red, 0; green, 0; blue, 0 }  ][line width=0.75]    (10.93,-3.29) .. controls (6.95,-1.4) and (3.31,-0.3) .. (0,0) .. controls (3.31,0.3) and (6.95,1.4) .. (10.93,3.29)   ;
\draw    (310,130) -- (222,130) ;
\draw [shift={(220,130)}, rotate = 360] [color={rgb, 255:red, 0; green, 0; blue, 0 }  ][line width=0.75]    (10.93,-3.29) .. controls (6.95,-1.4) and (3.31,-0.3) .. (0,0) .. controls (3.31,0.3) and (6.95,1.4) .. (10.93,3.29)   ;

\draw (214,205.4) node [anchor=north west][inner sep=0.75pt]  [font=\small]  {$0$};
\draw (391,205.4) node [anchor=north west][inner sep=0.75pt]  [font=\small]  {$L=a( N-1)$};
\draw (91,203.4) node [anchor=north west][inner sep=0.75pt]  [font=\small]  {$-aN_{\mathrm{wings}}$};
\draw (510,203.4) node [anchor=north west][inner sep=0.75pt]  [font=\small]  {$L+aN_{\mathrm{wings}}$};
\draw (322,153.4) node [anchor=north west][inner sep=0.75pt]    {$m$};
\draw (137,153.4) node [anchor=north west][inner sep=0.75pt]    {$m_{0} e^{\pm i\Delta \gamma ^{5}}$};
\draw (313,120) node [anchor=north west][inner sep=0.75pt]  [font=\large] [align=left] {bulk};
\draw (147,120) node [anchor=north west][inner sep=0.75pt]  [font=\large] [align=left] {wings};
\draw (467,120) node [anchor=north west][inner sep=0.75pt]  [font=\large] [align=left] {wings};
\draw (544,183.4) node [anchor=north west][inner sep=0.75pt]    {$x$};
\draw (457,153.4) node [anchor=north west][inner sep=0.75pt]    {$m_{0} e^{\mp i\Delta \gamma ^{5}}$};

\end{tikzpicture}
\caption{The schematic picture of the lattice setup with the wings regime for the iso-triplet state calculation at $0 \leq \theta <\pi$. 
We assign $m_0 e^{+i\Delta \gamma^5}$ ($m_0 e^{-i\Delta\gamma^5}$) on the left side of the wings 
and $m_0 e^{-i\Delta\gamma^5}$ ($m_0 e^{+i\Delta\gamma^5}$) on the right side for the first (second) flavor.}
\label{fig:wings-pion}
\end{figure}
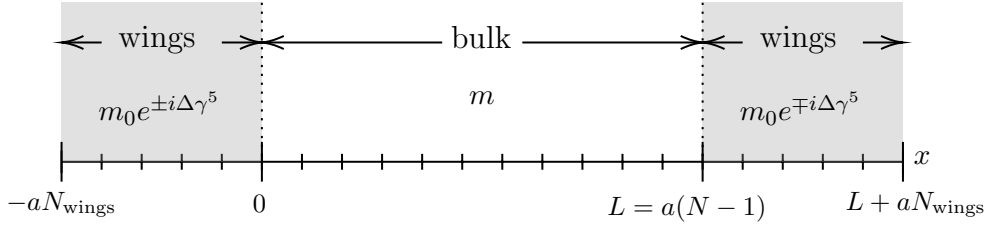
Thus, we take 
\begin{equation}
m_{\mathrm{wings}}=\begin{cases}
m_{0}\, e^{+i\Delta\gamma^{5}} & f=1,\\
m_{0}\, e^{-i\Delta\gamma^{5}} & f=2,
\end{cases}
\label{eq:twisted_mass}
\end{equation}
as the mass parameters in the wings.
Since the twisted mass breaks the isospin symmetry, 
the boundary between the wings and bulk can be a source of the triplet meson.
Thus, the one-point function of the pion becomes nonzero 
even in the trivially gapped phase $-\pi\leq\theta\leq\pi$ by setting $\Delta\neq 0$.
Note that there is a choice between having the same value of $\Delta$ on both sides or not. 
Here, we take the opposite signs of $\Delta$ on both sides of the wings as shown in figure~\ref{fig:wings-pion} 
to assign the opposite directions of the chiral rotation on each side.

Here, we address how the chiral rotation affects the lattice Hamiltonian.
In fact, the insertion of the flavor-asymmetric chiral rotation \eqref{eq:twisted_mass} also changes the fermion kinetic term, 
eq.~(\ref{eq:H_w}), of the lattice Hamiltonian.
In the wings regime, it is modified as follows: 
\begin{equation}
H_{w}=-i\sum_{f}\sum_{n \in N_{\rm wings}}\left(w-(-1)^{n}\frac{m_0}{2}\sin\Delta\right)\left(\chi_{f,n}^{\dagger}\chi_{f,n+1}-\chi_{f,n+1}^{\dagger}\chi_{f,n}\right).
\end{equation}
As for the mass term, eq.~\eqref{eq:H_m}, it is given by 
\begin{equation}
H_{m}= \sum_{f}\sum_{n \in N_{\rm wings}} \left(m_0\cos\Delta-\frac{N_{f}g^{2}a}{8}\right)(-1)^{n}\chi_{f,n}^{\dagger}\chi_{f,n},
\end{equation}
taking the $O(a)$ correction (\ref{eq:latticemass}) and the chiral rotation into account.

\subsubsection{Improvement on the fitting function}

The last step of the calculation is to fit the data of the one-point function and extract the mass.
In our previous work for $\theta = 0$, we fit the data points of $\log|\Braket{\mathcal{O}(x)}|$ by $-Mx+C$ to obtain the meson mass $M$. 
However, the boundary state $|\mathrm{Bdry}\rangle$ generically contains excited states such as two-particle states, which can give contamination. 
At $\theta \neq 0$, the mass gap is expected to become smaller, and thus the elimination of the excited-state contributions becomes more important. 

To incorporate such a contribution, we change the fitting ansatz.
We assume that the one-point functions behave as
\begin{equation}
\Braket{\mathcal{O}(x)}\sim Ae^{-Mx}+Be^{-(M+\Delta M)x},
\label{eq:1pt_ansatz}
\end{equation}
which is motivated by including the second-lowest state with a mass gap $M+\Delta M$.\footnote{
Although the contribution of the two-particle states can be Yukawa-type in general 
because of the integration over the relative momentum, 
the exponential form turned out to be enough for the practical analysis of our system.}
Then the effective mass, namely the logarithmic derivative of $\Braket{\mathcal{O}(x)}$, is given by 
\begin{equation}
-\frac{d}{dx}\log\Braket{\mathcal{O}(x)}\sim M+\frac{\Delta M}{1+Ce^{\Delta Mx}}.
\label{eq:Meff_1pt_ansatz}
\end{equation}
On the discretized lattice, we estimate the effective mass from the one-site difference of $\Braket{\mathcal{O}(x)}$,
\begin{equation}
-\frac{1}{a}\log\frac{\Braket{\mathcal{O}(x+a)}}{\Braket{\mathcal{O}(x)}}.
\label{eq:Meff_1pt_lat}
\end{equation}
Here, we also take the three-site average of the effective mass to suppress the fluctuation 
due to the discretization with the staggered fermion.

\subsection{Dispersion-relation scheme}

The dispersion-relation scheme examines the energy eigenstates themselves 
and can apply to the case of $\theta\neq 0$ straightforwardly.
Even though the parity and $G$-parity are not quantum numbers of the mesons, 
we can still identify the pion and sigma meson as the lowest-energy states 
with the isospin $J=1$ and $J=0$, respectively.
On the other hand, the eta meson becomes unstable at $\theta\neq 0$ as mentioned 
in section~\ref{subsec:spectrum_bosonization}, and thus it should disappear from the spectrum.

To make the dispersion relations more precise, 
we consider a technical improvement in DMRG for excited states.
In the DMRG with the Hamiltonian (\ref{eq:H_excited_states}), 
the sigma meson is relatively difficult to obtain compared with the pion.
The singlet states tend to appear at higher levels than the triplets 
since there are many momentum excitations of the pion triplets at low levels.
However, the higher states generally require more computational cost 
and suffer from larger systematic errors due to the lack of orthogonalities.

We deal with this problem by modifying the Hamiltonian.
The point is to generate the singlet states separately by the singlet projection, 
eliminating the states with $J>0$ from the low-energy spectrum.
For this purpose, we include the isospin Casimir operator $\bm{J}^{2}$ 
in the Hamiltonian as the cost term, 
\begin{equation}
H_{\ell}=H+W\sum_{\ell^{\prime}=0}^{\ell-1}\ket{\Psi_{\ell^{\prime}}}\bra{\Psi_{\ell^{\prime}}}+W_{J}\bm{J}^{2},
\label{eq:H_singlet}
\end{equation}
where $W_{J}>0$ is a parameter to be tuned. 
Since $\bm{J}^{2}$ is a positive-semidefinite operator, 
the last term increases the energy of the states with nonzero isospin $J>0$.
Then the condition $\bm{J}^{2}\ket{\Psi_{\ell}}=0$ is automatically imposed 
on the target state $\ket{\Psi_{\ell}}$ when minimizing the energy.
Since only the singlet states are involved in DMRG, the computational cost is reduced, 
which enables us to increase the number of sweeps to improve orthogonality.

\section{Simulation results}
\label{sec:result}

\subsection{Simulation parameters}

Let us explain the parameter setup in our calculation. 
Since the gauge coupling $g$ has mass dimension $1$, we measure the energy scale in the unit of $g$ by setting $g=1$. 
In this work, we always set the fermion mass $m=0.1$, so the photon mass is $\mu=\sqrt{2/\pi}\simeq 0.8$. 
The lattice size $N$, the lattice spacing $a$, a control parameter $\varepsilon$ of the truncation error in SVD, 
and the number of iterations $N_{\mathrm{sweep}}$ in DMRG are summarized in table~\ref{tab:summary_parameter}~\footnote{
The explicit value of $a$ for the correlation-function and one-point-function schemes is given by solving $a\times(160-1)=0.2\times(200-1)$.}.
We use the C\texttt{++} library of ITensor~\cite{itensor} in our calculation.

\begin{table}[htb]
    \centering
    \begin{tabular}{c|ccc|cc}
        scheme               & $N$   & $a$    & $L=(N-1)a$ & $\varepsilon$          & $N_{\mathrm{sweep}}$\\ \hline
        correlation function & $160$ & $0.25$ & $39.8$     & $10^{-10,-12,-14,-16}$ & $20$ \\
        one-point function   & $320$ & $0.25$ & $79.9$     & $10^{-10}$             & $20$ \\
        dispersion relation  & $100$ & $0.20$ & $19.8$     & $10^{-10}$             & $50$, $80$ \\
    \end{tabular}
    \caption{Summary of the parameter setup for each calculation step. $L$ is the physical lattice volume in the unit $g=1$.}
    \label{tab:summary_parameter}
\end{table}

In the calculation of the correlation function, we take a middle lattice size $N=160$. 
In our previous simulation, the data of the correlation function strongly depend on the truncation parameter $\varepsilon$, 
thus we take several values of $\varepsilon$ and estimate the cutoff dependence.
In the measurement of the one-point functions, the cutoff parameter is fixed to $\varepsilon=10^{-10}$, 
since we found that its cutoff dependence is negligible. 
Instead, we take a large lattice size $N=320$ to see the long-distance regime from the boundary.

To generate the ground and excited states for the dispersion-relation scheme, we take a small lattice site $N=100$.
We set $N_{\rm sweep}=50$ when we use the cost Hamiltonian~(\ref{eq:H_excited_states}), 
where all eigenstates of the original Hamiltonian are generated. 
On the other hand, we set $N_{\rm sweep}=80$ when we use the cost Hamiltonian~(\ref{eq:H_singlet}), 
where only singlet states are generated.
Thanks to the focus on singlet only, the number of levels to be calculated is reduced in the latter case. 
Then we can increase the number of sweeps while keeping a reasonable calculation cost~\footnote{
As we go to the higher level, the bond dimension tends to increase.
Thus, the total computational cost is not linear on the number of the target states but higher.}.
The other technical simulation parameters for each scheme will be given later.

\subsection{Improved one-point-function scheme for \texorpdfstring{$0 \leq \theta < \pi$}{0<=theta<pi}}
\label{subsec:improved_1pt_scheme}

Now, we present the results of the improved one-point-function scheme.
Note that we obtain the mass spectra of the pion and sigma meson in the range of $0 \leq \theta < \pi$.
As for the eta meson, it becomes unstable in a large $\theta$ regime.
Thus, we put the analyses of the eta meson in appedix~\ref{sec:fate_of_eta} instead of the main text.
As we show, at $\theta=\pi$, the data for the one-point function for both pion and sigma meson do not fit the ansatz~(\ref{eq:Meff_1pt_ansatz}) 
since the system becomes almost gapless and CFT-like there.
We have a detailed discussion about this point in section~\ref{sec:Result_CFT} independently.

\subsubsection{Determination of the mixing angle}
\label{subsec:correlation_matrix}

First, we resolve the mixing of the operator using the two-point correlation function.
As explained in sections~\ref{subsec:theory_at_critical_point} and~\ref{sec:def-mixing-matrix}, 
the axial rotation by $\delta_{-}=\theta/2$ is expected for the pion while the extra rotation, 
$\omega(\theta)=O((m/g)^{4/3}\sin(\theta/2)(\cos(\theta/2))^{1/3})$, 
is required for the sigma and eta mesons, hence, $\delta_{+}=\theta/2+\omega(\theta)$.

\begin{figure}[htb]
\centering
\includegraphics[scale=0.5]{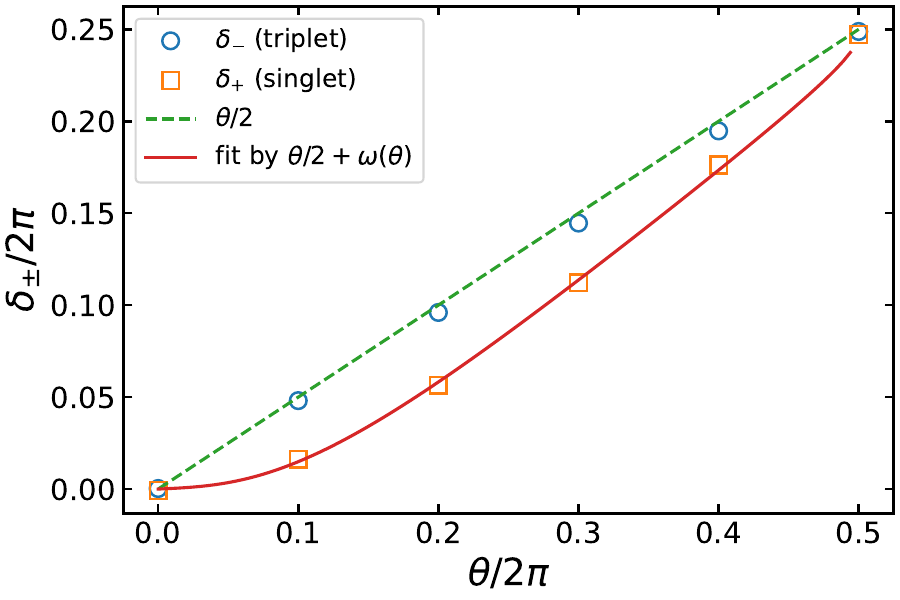} 
\caption{\label{fig:delta_vs_theta}
The mixing angles $\delta_{-}$ for the triplet and $\delta_{+}$ for the singlet sector are plotted against $\theta/2\pi$.
The dashed line denotes $\theta/2$ for comparison with $\delta_{-}$.
The result of fitting $\delta_{+}$ by $\theta/2 + \omega(\theta)$ is plotted as the solid curve.}
\end{figure}

We plot the numerical results of the mixing angles $\delta_{\pm}$ as a function of $\theta/2\pi$ in figure~\ref{fig:delta_vs_theta}.
To obtain this, we use the correlation matrix $\boldsymbol{C}_{\pm}(x,y)$ for $x=(L-r)/2$ and $y=(L+r)/2$ 
with the distance $r=15$ in our simulation. 
The angle $\delta_-$ for the iso-triplet sector is shown with the empty blue circles and mostly agrees with the line of $\theta/2$ as expected.
We suspect that the small deviation comes from the contributions of the excited state such as the two-pion states. 
The angle $\delta_+$ for the iso-singlet sector is shown with the empty orange squares, which have a clear deviation from $\theta/2$. 
We fit the numerical result by $\theta/2+\omega(\theta)$ using eq.~\eqref{eq:M-theta} with parameters $A$ and $B$ 
and obtain the best-fit values $A=-0.23(2)$ and $B=0.76(4)$.
The fitting curve is shown as the solid curve in the figure.
The success of this fitting strongly indicates the consistency with analytic predictions on the $\eta$-$\sigma$ mixing at finite $m$.

\subsubsection{Sigma meson masses}
\label{subsec:1pt_sigma}

Now, let us show the measurement results of the one-point function for the sigma meson and extract the mass. 
In the following analyses, 
we set the number of lattice sites in the wings to $N_{\mathrm{wings}}=20$.
The total number of the site for the staggered fermion is then $N+2N_{\mathrm{wings}}=360$.
We set the fermion mass in the wings to $m_\mathrm{wings}=m_{0}=10$, which is a hundred times larger than the mass $m=0.1$ in the bulk.
We generate the ground states of the whole system including the wings 
by DMRG for $0 \leq \theta/2\pi  \leq 0.5$ with the interval $\Delta (\theta/2\pi) =0.1$.
Since attaching the wings does not change the bound dimension of MPS so much, the increase in the computational cost is negligible.

\begin{figure}[htb]
\centering
\includegraphics[scale=0.45]{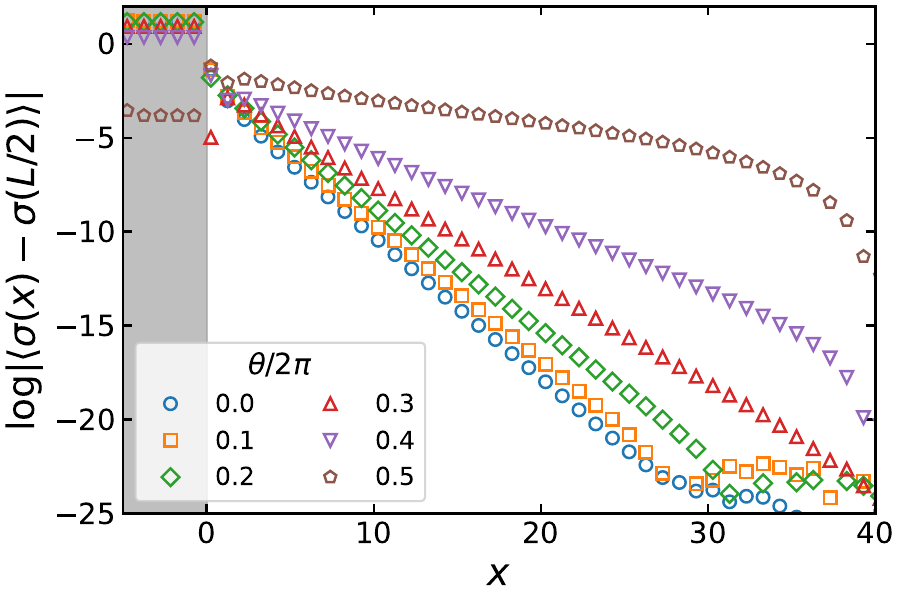}
\includegraphics[scale=0.45]{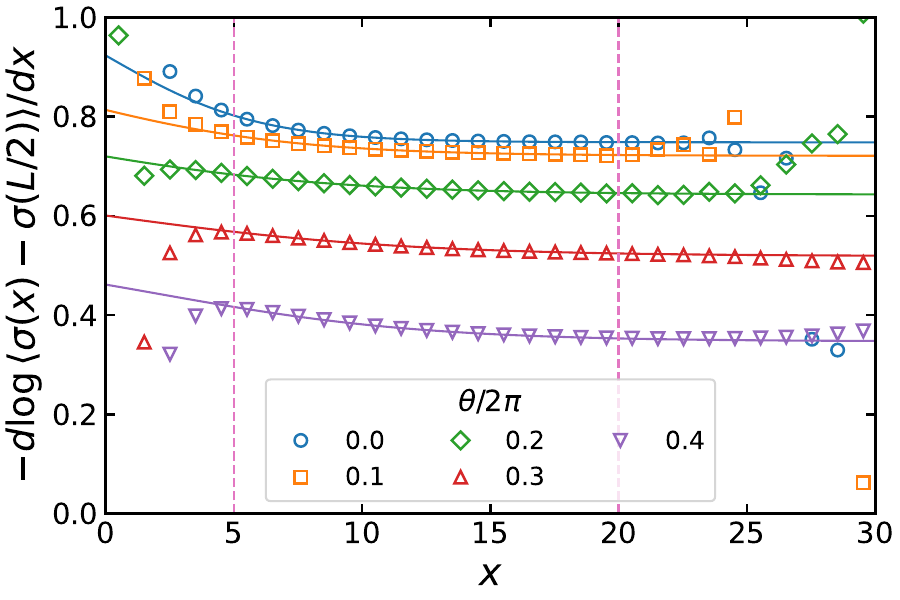} 
\caption{\label{fig:1pte_sigma}
(Left) The one-point function $\log|\Braket{\sigma(x)-\sigma(L/2)}|$ of the sigma meson is plotted 
against the distance $x$ from the boundary.
The value at $x=L/2$ is subtracted from $\Braket{\sigma(x)}$ to eliminate the constant shift in the bulk.
The shaded region indicates the wings.
(Right) The effective mass calculated from the one-point function in the left panel is plotted against $x$.
The results of fitting by (\ref{eq:Meff_1pt_ansatz}) are shown as well.
The fitting range is between the vertical dashed lines.}
\end{figure}

The results for the one-point function $\Braket{\sigma(x)}$ of the sigma meson operator (\ref{eq:meson_op}) 
for the ground states are shown in the left panel of figure~\ref{fig:1pte_sigma}.
From these data of the one-point function, we compute the effective mass and fit the data using the improved fitting function 
eq.~(\ref{eq:Meff_1pt_ansatz}) in the range $5\leq x\leq 20$ with parameters $M$, $\Delta M$, and $C$. 
The results are shown in the right panel of figure~\ref{fig:1pte_sigma}, 
and the fitting results of the parameters are summarized in table~\ref{tab:fit_1pte_sigma}.
\begin{table}[htb]
\centering
\begin{tabular}{|c|c|c|c|}\hline $\theta/2\pi$ & $M_{\sigma}$ & $\Delta M$ & $C$ \tabularnewline\hline \hline 0.0 & 0.74814(9) & 0.344(3) & 0.173(3) \tabularnewline\hline 0.1 & 0.7209(3) & 0.227(5) & 0.47(1) \tabularnewline\hline 0.2 & 0.6429(2) & 0.188(2) & 0.563(3) \tabularnewline\hline 0.3 & 0.5188(2) & 0.167(1) & 0.453(2) \tabularnewline\hline 0.4 & 0.3471(4) & 0.192(2) & 0.259(2) \tabularnewline\hline \end{tabular}
\vspace{-1.3em}
\caption{\label{tab:fit_1pte_sigma}
The fitting results of the effective mass
computed from the one-point function $\Braket{\sigma(x)}$ of the sigma meson.
The errors of these values come from the fitting error.}
\end{table}
Note that in the plot and table, we show the data except for $\theta = \pi$ since the fit ansatz, eq.~(\ref{eq:Meff_1pt_ansatz}), 
does not work at this $\theta$ where it closes to the (nearly) conformal theory.
We can see that the effective mass is not completely flat but slightly curved in the bulk. 
It indicates that the correction term from the excited state is not negligible~\footnote{
In our previous work~\cite{Itou:2023img}, we observed small differences between the results obtained by the three methods at $\theta=0$.
A part of the differences can be explained by neglecting the contribution from the excited state in the fitting of the one-point function.}.

\subsubsection{Pion masses}
\label{subsec:1pt_pi}

Next, let us focus on the iso-triplet meson, namely the pion.
We generate the ground state with the twisted fermion mass assigned to the wings region 
by DMRG for $\theta/2\pi=0.0,0.1,\cdots,0.5$.
As for the wings regime, we set the twist angle $\Delta=0.1$, the mass $m_{0}=10$, and the size $N_{\mathrm{wings}}=20$ 
in figure~\ref{fig:wings-pion}.

\begin{figure}[htb]
\centering
\includegraphics[scale=0.45]{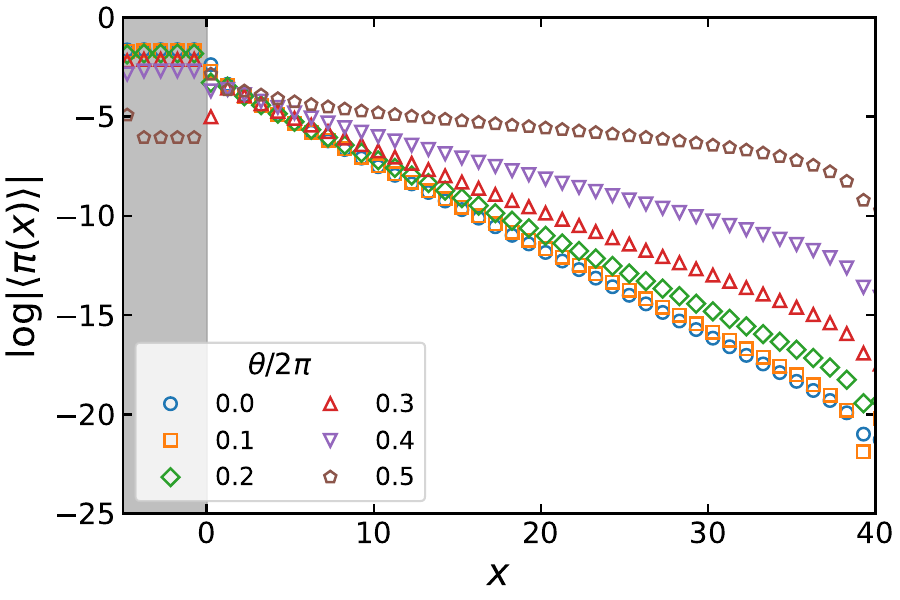}
\includegraphics[scale=0.45]{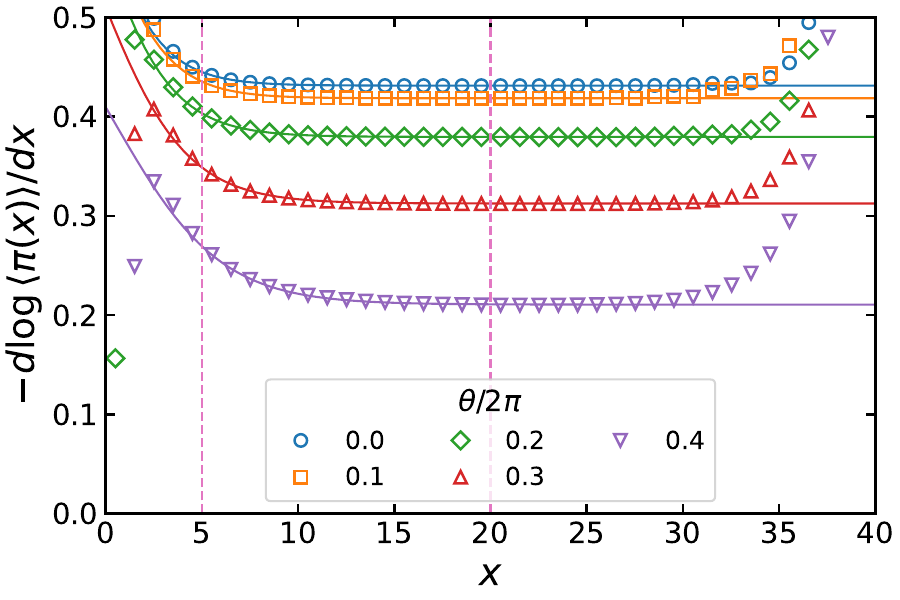}
\caption{\label{fig:1pte_pi}
(Left) The one-point function $\log|\Braket{\pi(x)}|$ of the pion 
is plotted against the distance $x$ from the boundary.
The shaded region indicates the wings.
(Right) The effective mass calculated from the one-point function in the left panel is plotted against $x$. 
The results of fitting by (\ref{eq:Meff_1pt_ansatz}) are shown as well.
The fitting range is between the vertical dashed lines.}
\end{figure}

We measure the one-point function $\Braket{\pi(x)}$ of the pion (\ref{eq:meson_op}), 
where the rotation matrix $R_{-}$ is determined by the result in section~\ref{subsec:correlation_matrix}.
The results of the one-point function and the corresponding effective mass are shown in figure~\ref{fig:1pte_pi}.
We obtain the nonzero one-point function thanks to the twisted mass assigned in the wings regime.
To obtain the effective mass by fitting with the ansatz eq.~(\ref{eq:Meff_1pt_ansatz}), 
we utilize the data in the range $5\leq x\leq 20$ as shown in the right panel of figure~\ref{fig:1pte_pi}.
The best-fit values of the parameters are summarized in table~\ref{tab:fit_1pte_pi}.
\begin{table}[htb]
\centering
\begin{tabular}{|c|c|c|c|}\hline $\theta/2\pi$ & $M_{\pi}$ & $\Delta M$ & $C$ \tabularnewline\hline \hline 0.0 & 0.431205(8) & 0.562(2) & 0.145(3) \tabularnewline\hline 0.1 & 0.418442(8) & 0.541(2) & 0.139(2) \tabularnewline\hline 0.2 & 0.37955(1) & 0.498(2) & 0.137(2) \tabularnewline\hline 0.3 & 0.31242(3) & 0.441(2) & 0.135(2) \tabularnewline\hline 0.4 & 0.2106(1) & 0.366(3) & 0.135(3) \tabularnewline\hline \end{tabular}
\vspace{-1.3em}
\caption{\label{tab:fit_1pte_pi}
The fitting results of the effective mass
computed from the one-point function $\Braket{\pi(x)}$ of the pion.
The errors of these values come from the fitting error.}
\end{table}
Note that the fitting ansatz~(\ref{eq:Meff_1pt_ansatz}) does not work at $\theta=\pi$ again 
and we will discuss its CFT-like behavior in section~\ref{sec:Result_CFT}.

\subsubsection{Summary of the one-point-function scheme}

\begin{figure}[htb]
\centering
\includegraphics[scale=0.5]{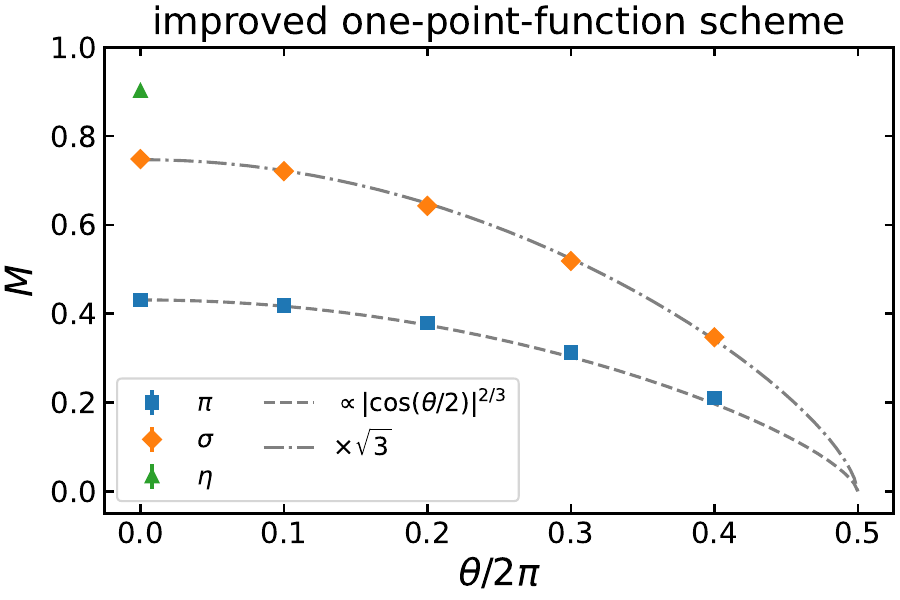}
\caption{\label{fig:M_all_1pte}
The masses of the pion and sigma meson obtained by the one-point-function scheme 
with the wings are plotted against $\theta/2\pi$. 
The eta-meson mass at $\theta=0$ is also plotted for reference.}
\end{figure}

Let us summarize the results of the one-point-function scheme.
The $\theta$-dependent masses of the pion and sigma meson are shown in figure~\ref{fig:M_all_1pte}.
The mass of the eta meson at $\theta=0$ is also plotted for reference.
The gray dashed curve denotes $M_{\pi}(0)|\cos(\theta/2)|^{2/3}$, 
where the overall coefficient $M_{\pi}(0)$ is determined by the numerical result at $\theta=0$.
The $\theta$-dependence of the pion mass is consistent 
with the analytic calculation by the bosonized model in eq.~\eqref{eq:M_pi_theta}.
Indeed, we can see that the pion mass decreases as $\theta$ increases, approaching the almost gapless phase at $\theta=\pi$.
Furthermore, the gray dash-dot curve denotes $\sqrt{3}M_{\pi}(0)|\cos(\theta/2)|^{2/3}$.
The result of the sigma-meson mass is also consistent with the result of the WKB approximation in eq.~\eqref{eq:sqrt3-Schwinger}.
Our result indicates that this relation holds not only around $\theta=0$ but also in the large $\theta$ region.

\subsection{Dispersion-relation scheme}

Now, let us discuss the second calculation scheme, namely the dispersion-relation scheme. 
We first generate the eigenstates up to the level $\ell=30$ using the cost Hamiltonian~\eqref{eq:H_excited_states}, 
where we set $W=10$ in our calculations.
For each state of the level $\ell$, 
we measure the energy gap $\Delta E_{\ell}=E_{\ell}-E_{0}$ 
and the momentum square $\Delta K_{\ell}^2=\Braket{K^2}_{\ell}-\Braket{K^2}_{0}$, 
where the result $\Braket{K^2}_{0}$ of the ground state is subtracted 
from $\Braket{K^2}_{\ell}$ to remove the nonzero offset.

\begin{figure}[htb]
\centering
\includegraphics[scale=0.35]{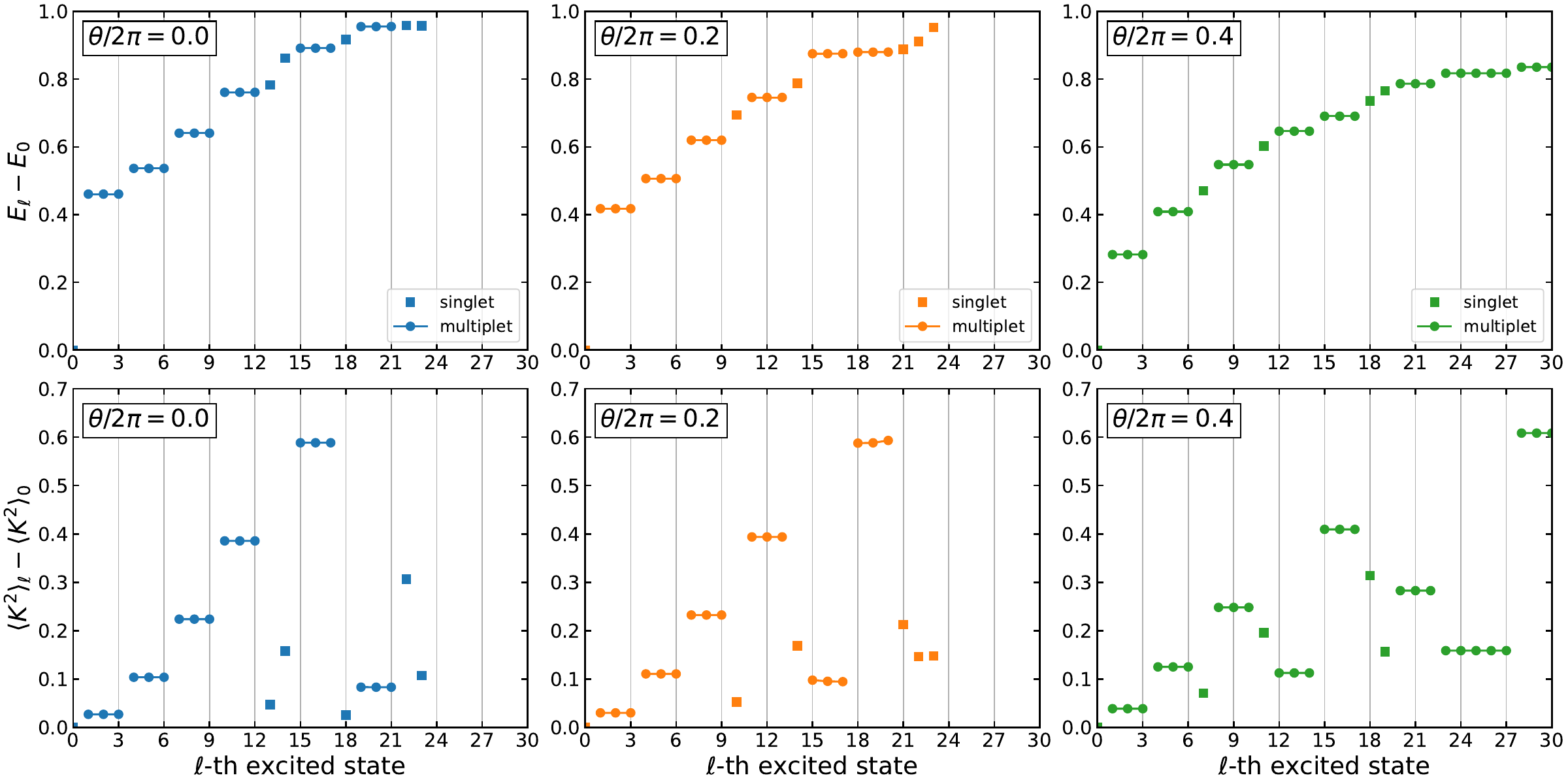} 
\caption{\label{fig:E_and_K2}
(Top) The energy gap $\Delta E_{\ell}=E_{\ell}-E_{0}$
is plotted against the level $\ell$ of the excited state.
(Bottom) The square of total momentum 
$\Delta K_{\ell}^{2}=\Braket{K^{2}}_{\ell}-\Braket{K^{2}}_{0}$
is plotted against $\ell$ after subtracting the result of the ground state.
The left, center, and right columns correspond to 
$\theta/2\pi=0.0$, $0.2$, and $0.4$, respectively. 
Some higher states are omitted due to insufficient convergence.}
\end{figure}

The results of $\Delta E_{\ell}$ and $\Delta K_{\ell}^{2}$ 
for $\theta/2\pi=0.0,0.2,0.4$ are shown in figure~\ref{fig:E_and_K2}.
We first identify the series of triplets with the monotonically increasing momentum 
$\Delta K_{\ell}^{2}$ starting from the lowest triplet $\ell=1,2,3$ as the pions. 
Indeed, they have the isospin quantum numbers, $J=1$ and $J_{z}=0,\pm1$, 
which indicates that they are consistent with the quantum numbers of the pions.
On the other hand, there is another series of $J=1$ triplets at higher energy levels, 
which is expected to be the scattering states investigated in ref.~\cite{Harada:1993va}.
Furthermore, interestingly, we also find quintet states, 
namely 5-fold degenerated states for $\theta/2\pi\geq 0.3$, for instance, $\ell=23,\cdots,27$ 
in the right panel of figure~\ref{fig:E_and_K2}.
We found that they have $\bm{J}^{2}=6$ and $J_{z}=0,\pm1,\pm2$.
Thus, they are the two-pion states with $J=2$, which comes from the combination of the two pions with $J=1$.

Next, we generate only the singlet excited states up to the level $\ell=10$ 
by DMRG with the Hamiltonian~\eqref{eq:H_singlet}.
The parameters are set to $W=10$ and $W_{J}=1$ so that DMRG could have a good convergence.
Here we perform more sweeps, $N_\mathrm{sweep}=80$, to improve the precision as shown in table~\ref{tab:summary_parameter}.

\begin{figure}[htb]
\centering
\includegraphics[scale=0.35]{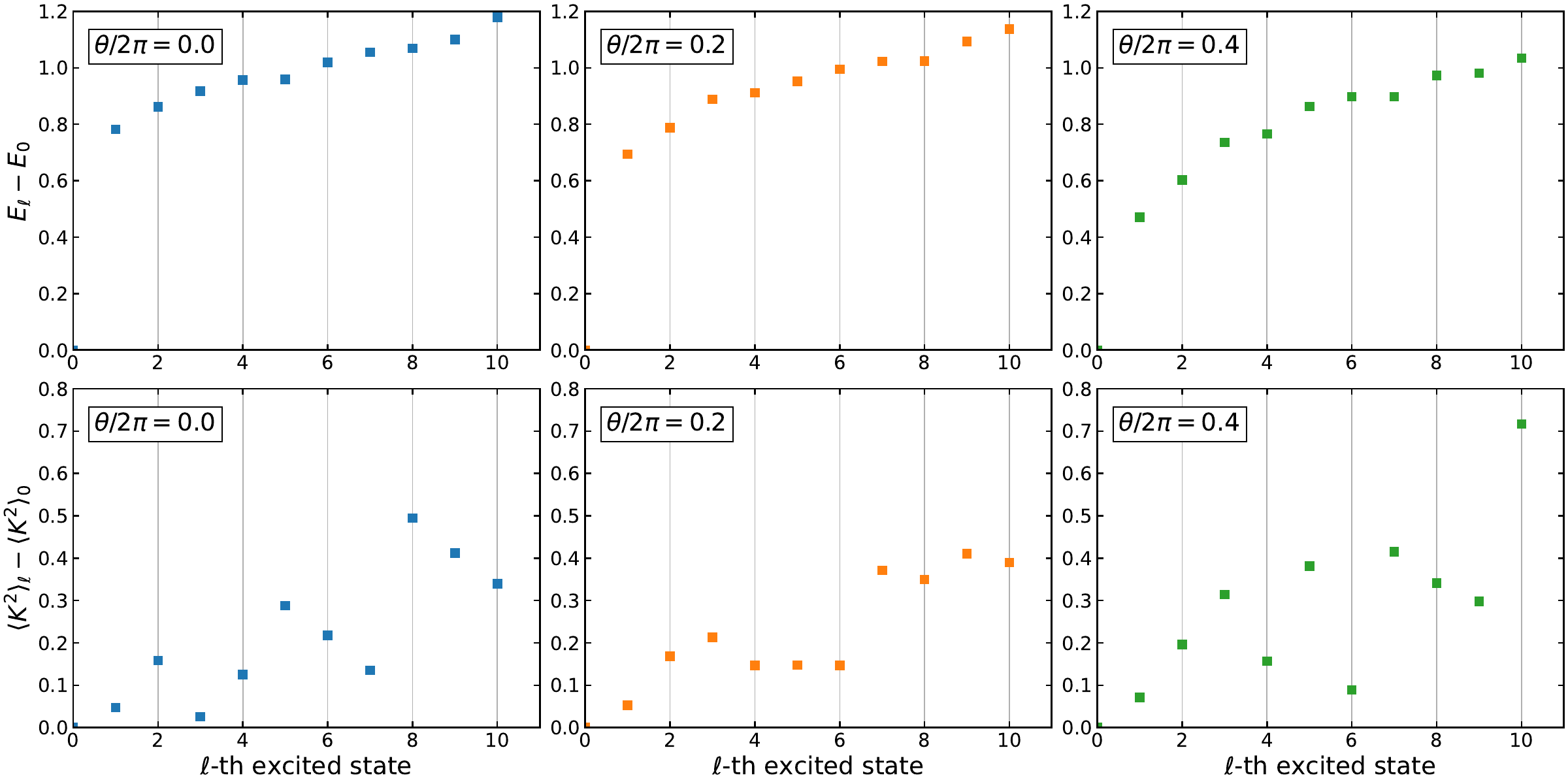} 
\caption{\label{fig:E_and_K2_J2}
These results are obtained by DMRG with the singlet projection~\eqref{eq:H_singlet}.
(Top) The energy gap $\Delta E_{\ell}=E_{\ell}-E_{0}$
is plotted against the level $\ell$ of the excited state. 
(Bottom) The square of total momentum 
$\Delta K_{\ell}^{2}=\Braket{K^{2}}_{\ell}-\Braket{K^{2}}_{0}$
is plotted against $\ell$ after subtracting the result of the ground state. 
The left, center, and right columns correspond to 
$\theta/2\pi=0.0$, $0.2$, and $0.4$, respectively.}
\end{figure}

The results of the energy gap $\Delta E_{\ell}$ 
and the momentum square $\Delta K_{\ell}^{2}$ for $\theta/2\pi=0.0,0.2,0.4$ 
are shown in figure~\ref{fig:E_and_K2_J2}. 
We first confirm that the states obtained by the Hamiltonian~\eqref{eq:H_singlet} 
have the isospin quantum numbers consistent with $\bm{J}^{2}=0$ and $J_{z}=0$. 
Furthermore, $\Delta E_{\ell}$ and $\Delta K_{\ell}^{2}$ in figure~\ref{fig:E_and_K2_J2} 
are almost consistent with those in figure~\ref{fig:E_and_K2}, while some data show small discrepancy.
The discrepancy appears in the states, which have almost degenerate energies with other states in figure~\ref{fig:E_and_K2}. 
We expect that the results obtained by the singlet projection are more reliable 
since they can preserve the orthogonal condition without accidental degeneracy.
Therefore, to obtain the dispersion relation of the sigma and eta mesons, 
we use the results of the singlet projection in the following.

Now, let us focus on the results of the singlet projection in figure~\ref{fig:E_and_K2_J2}.
We first identify the lowest series of singlet states 
with the monotonically increasing momentum as the sigma meson. 
At $\theta=0$, we can confirm that the corresponding state has the positive $G$-parity, namely $\Braket{G}>0$.
We also find the singlet states with $\Braket{G}<0$, namely the eta meson, at $\theta=0$. 
For $\theta\neq 0$, the eta meson is no longer a stable particle, 
and thus the corresponding states might be replaced by some scattering state. 
We find a series of eta-meson-like states at $\theta/2\pi=0.1$ but not for $\theta/2\pi\geq0.2$. 
By gradually increasing $\theta$ in the range $0.1\leq\theta/2\pi\leq0.2$, 
we observe that a state with $\Braket{G}<0$ in the spectrum suddenly changes into a state with $\Braket{G}>0$.

\begin{figure}[htb]
\centering
\includegraphics[scale=0.35]{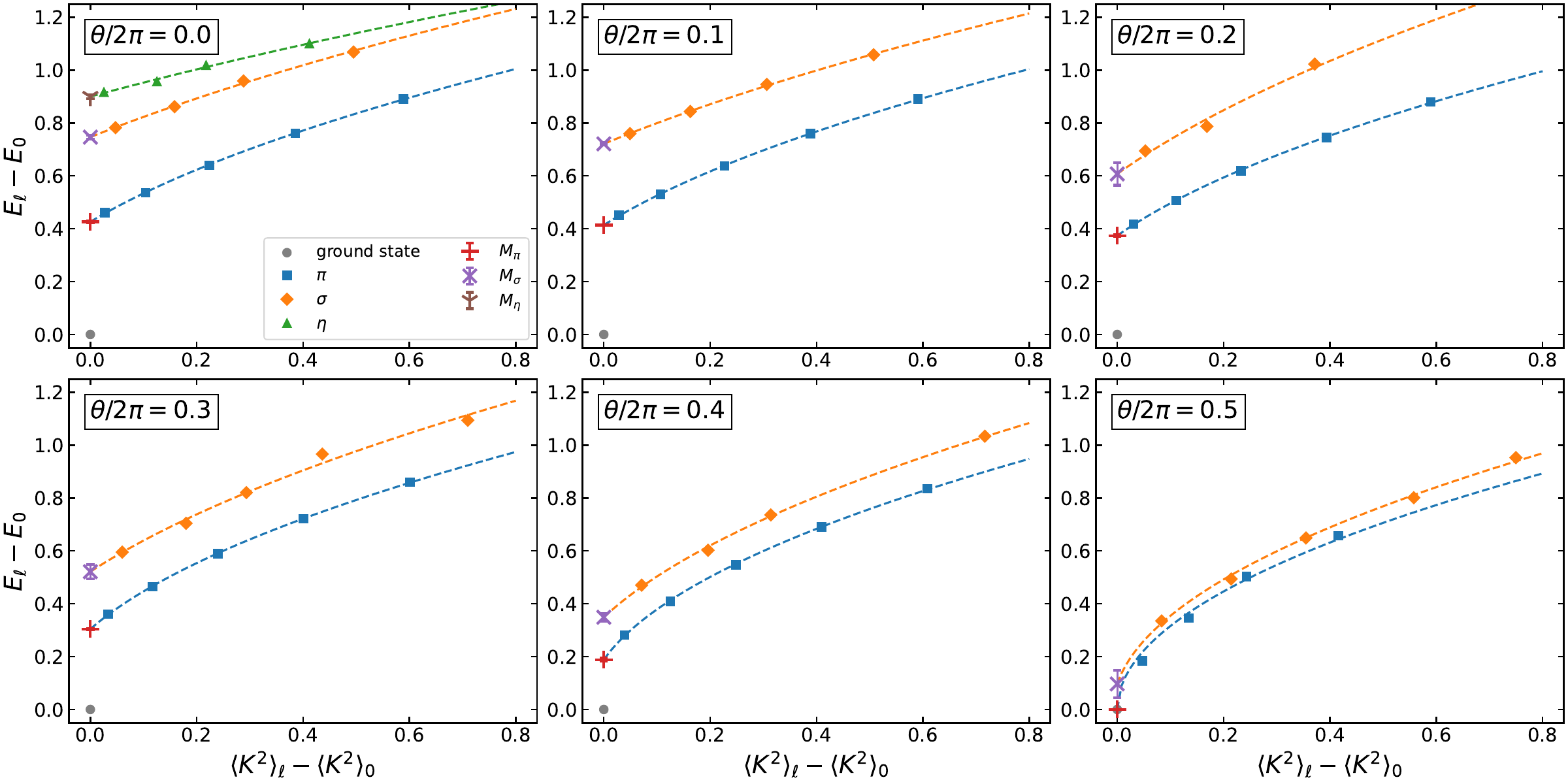}
\caption{\label{fig:E_vs_K2}
The energy gap $\Delta E_{\ell}$ is plotted against the square of total momentum 
$\Delta K_{\ell}^{2}$ for $\theta/2\pi=0.0,0.1,\cdots,0.5$.
The states identified as the same meson are denoted by the same symbol.
We fit the data points for each meson by $\Delta E=\sqrt{b^{2}\Delta K^{2}+M^{2}}$, 
and the results are shown by the broken curves.
The markers with error bars at the left endpoints denote the extrapolated value and its error from the fitting.}
\end{figure}

Finally, we investigate the dispersion relations using the results of the pion and sigma meson.
We plot the energy gap $\Delta E_{\ell}$ against the momentum square 
$\Delta K_{\ell}^{2}$ for each meson in figure~\ref{fig:E_vs_K2}.
The states identified as the same meson are plotted by the same symbol. 
Then we fit these data points by $\Delta E=\sqrt{b^{2}\Delta K^{2}+M^{2}}$ with fitting parameters $M$ and $b$. 
The fitting result of $M$ can be regarded as the mass of the corresponding meson 
as an extrapolation to $\Delta K^{2}\rightarrow 0$.
The parameter $b$ is included taking account of possible lattice artifacts on the momentum.
As for the pion at $\theta=\pi$, 
we fit the data points by $\Delta E= \sqrt{b^2 \Delta K^{2} }$ assuming $M=0$ 
because the fitting including the parameter $M$ is unstable 
due to the square root of a small number.\footnote{The error of $M$ is roughly $O(1/M)$ considering the error propagation.}

\begin{figure}[htb]
\centering
\includegraphics[scale=0.5]{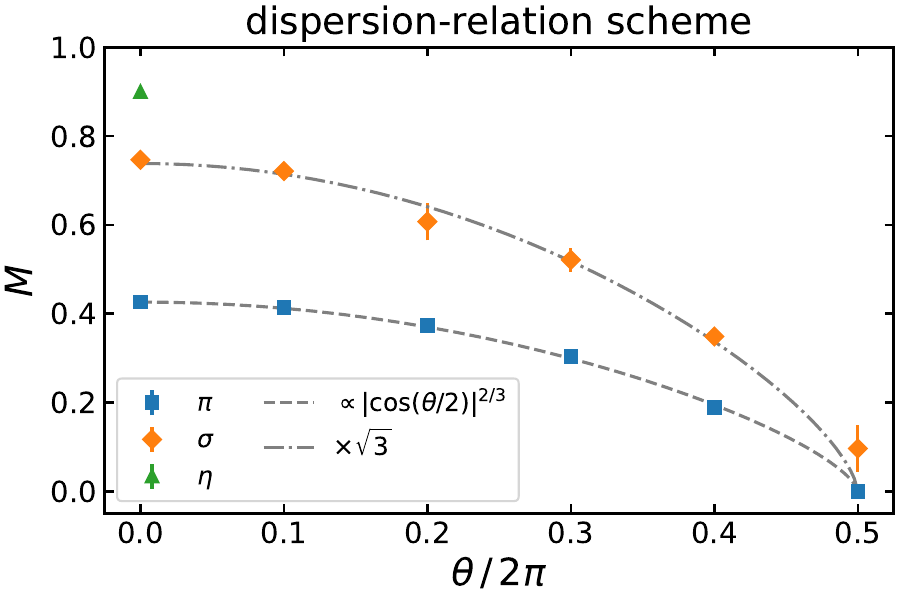} 
\caption{\label{fig:M_all_disp}
The masses of the pion, sigma, and eta meson obtained 
by the dispersion-relation scheme are plotted against $\theta/2\pi$.}
\end{figure}

The $\theta$-dependent masses of the pion, sigma, and eta meson 
are summarized in figure~\ref{fig:M_all_disp}. 
Now, we obtain the masses for the pion and sigma meson at $\theta=\pi$, which are almost zero as expected.
Again, the gray dashed curve denotes the calculation in the bosonized model, 
$M_{\pi}(0)|\cos(\theta/2)|^{2/3}$, where the overall coefficient $M_{\pi}(0)$ 
is determined by the numerical result at $\theta=0$. 
The gray dash-dot curve denotes $\sqrt{3}M_{\pi}(0)|\cos(\theta/2)|^{2/3}$.
Both the pion and sigma-meson masses almost agree with the analytic predictions.

Let us make some comments on the fate of the eta meson.
We find some states similar to the eta meson, 
but the energies of such states become much higher for $\theta/2\pi \geq 0.2$ than that for $\theta/2\pi \leq 0.1$.
In fact, the data points of $\Delta E_{\ell}$ and $\Delta K_{\ell}^{2}$ 
of these states are no longer on a smooth line of the dispersion relation.
Thus, it looks like the eta meson has disappeared from the spectrum, 
which is consistent with that the eta meson is no longer stable and decays into the pions or mixes with the sigma meson.
It also causes the relatively large error of the sigma-meson mass at $\theta/2\pi=0.2$.
Indeed, the spectrum is changed drastically around $\theta/2\pi = 0.2$.
It suggests that the sigma meson is strongly contaminated by mixing with the eta meson 
and they might be in the process of being exchanged and mixed.

\section{One-point function at \texorpdfstring{$\theta=\pi$}{theta=pi}}
\label{sec:Result_CFT}

The massive $2$-flavor Schwinger model at $\theta=\pi$ has the tiny mass gap $M_{\pi}\sim e^{-\# g^{2}/m^{2}} g$~\cite{Dempsey:2023gib}, 
and we expect almost conformal behavior. 
We here compare the behaviors of the one-point function with its analytic computation on a finite interval 
with the $SU(2)_1$ WZW conformal theory as shown in section~\ref{subsec:theory_at_critical_point}.

\begin{figure}[htb]
\centering
\includegraphics[scale=0.45]{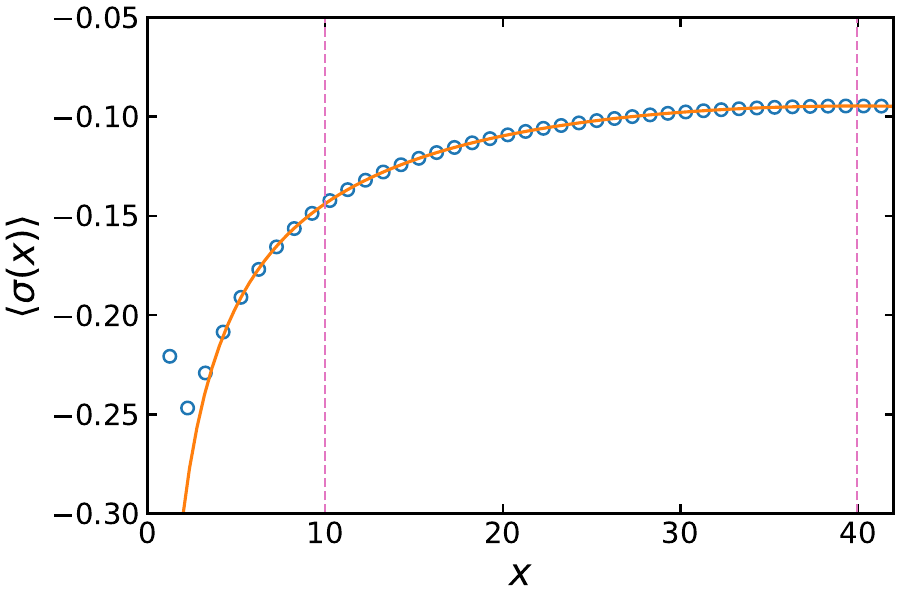}
\includegraphics[scale=0.45]{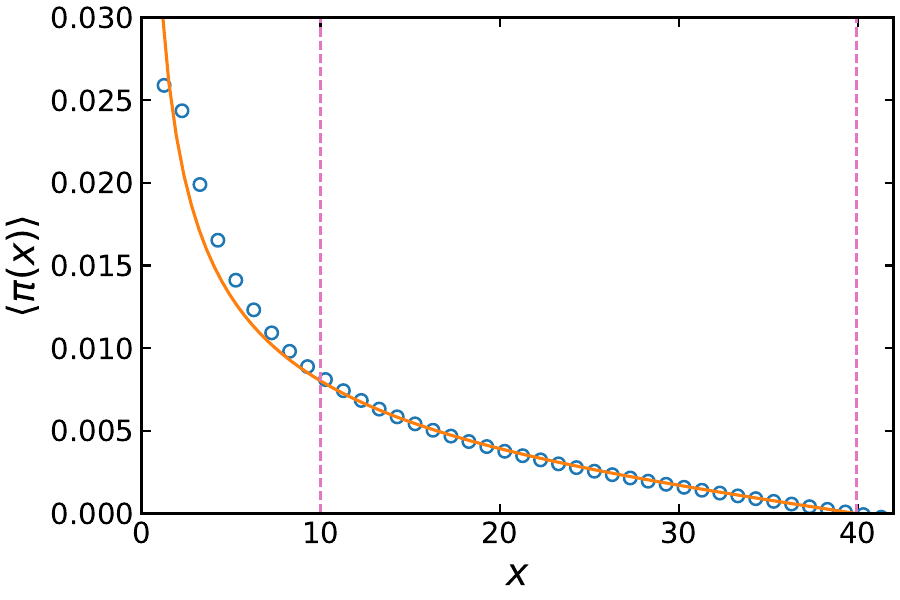}
\caption{\label{fig:cft_sigma}
The one-point functions at $\theta=\pi$ of the sigma meson $\Braket{\sigma(x)}$ (left) 
and the pion $\Braket{\pi(x)}$ (right) are plotted against the distance $x$ from the boundary. 
We use the lattice equipped with the wings regime, 
setting $N=320$ in the bulk and $N_{\mathrm{wings}}=20$ in the wings.
The fermion mass in the wings is $m_{\mathrm{wings}}=m_0$ for the sigma meson 
and $m_{\mathrm{wings}}=m_0 e^{\pm i\Delta\gamma^{5}}$ for the pion with $m_0=10$.
In the latter case, the twist parameters are $\Delta=+0.1$ and $-0.1$ 
on the left and right sides of the wings, respectively.
The fitting curves of $A/\sqrt{\sin(\pi x/L)}+B$ for the sigma meson 
and $A\sin\left[\Delta(1-2x/L)\right]/\sqrt{\sin(\pi x/L)}$ for the pion are shown as well, 
where the fitting range is between the vertical dashed lines.}
\end{figure}

For the sigma meson operator, the analytic result of the one-point function 
$\Braket{\sigma(x)}$ is given by eq.~\eqref{eq:cft_sigma}.
The corresponding numerical result obtained by DMRG at $\theta=\pi$ 
is shown in the left panel of figure~\ref{fig:cft_sigma}.
In this calculation, we use the lattice equipped with the wings regime, 
in which the large fermion mass $m_{\mathrm{wings}}=m_0=10$ is assigned as in section~\ref{subsec:1pt_sigma}.

We fit the data points of $\Braket{\sigma(x)}$ by $A/\sqrt{\sin(\pi x/L)}+B$ 
with the parameters $A$ and $B$. 
Then we obtain the best-fit values $A=-0.079922(8)$ and $B=-0.014741(9)$.
The corresponding fitting curve is plotted in the left panel of figure~\ref{fig:cft_sigma}. 
We find that the numerical result agrees with (\ref{eq:cft_sigma}) in the bulk region. 
The deviation around the boundary can be interpreted as the contribution of the eta meson 
and the artifact of the staggered fermion.\footnote{
The local operators of the Dirac fermion consist of the multi-point operators 
of the staggered fermion such as $\chi_{n}^{\dagger}\chi_{n+1}$.
Thus, the boundary condition for the Dirac fermion is not straightforwardly translated into that for the staggered fermion.
In particular, it violates the discrete chiral symmetry explicitly as it is realized as the one-unit lattice translation.}
We would like to note that the nonzero value of the constant part $B$ is quite suggestive 
as it can be regarded as the expectation value for the chiral condensate detecting the $(\mathbb{Z}_2)_{G+\mathrm{chiral}}$ symmetry breaking, 
while we need more systematic studies to make it conclusive. 

For the pion, we consider the lattice with the wings 
and take the large fermion mass with the flavor-asymmetric chiral rotation 
$m_{\mathrm{wings}}=m_0 e^{\pm i\Delta\gamma^{5}}$ in the wings regime as shown in figure~\ref{fig:wings-pion}.
Here, we take $m_0=10$ and $\Delta=0.1$.
The directions of the twist are opposite, so that we obtain eq.~\eqref{eq:cft_pi_odd} which has a nontrivial $\Delta$ dependence. 

The numerical result of the pion one-point functions $\Braket{\pi(x)}$ is displayed 
in the right panel of figure~\ref{fig:cft_sigma}.
We fit the data points by $A\sin\left[\Delta(1-2x/L)\right]/\sqrt{\sin(\pi x/L)}$ and obtain $A=0.0663(1)$.
As can be seen from the figure, the numerical results mostly agree with the analytic result, 
eq.~\eqref{eq:cft_pi_odd} in the bulk region.
To confirm the nontrivial $\Delta$ dependence, 
we also check that the one-point function behaves as eq.~\eqref{eq:cft_pi_odd}
for other choices of the twist parameter, $\Delta=0.2,0.3$.
Therefore, it is numerically confirmed that the pion and sigma meson 
in the $2$-flavor Schwinger model at $\theta=\pi$ are well described 
by SU(2) level-1 WZW CFT on a finite interval.

\section{Conclusion and discussion}
\label{sec:conclusion}

Extending our previous study~\cite{Itou:2023img}, 
we investigate the physics of the massive $2$-flavor Schwinger model at nonzero $\theta$ by employing the DMRG to its lattice Hamiltonian formalism. 
In ref.~\cite{Itou:2023img}, we have proposed three methods to obtain the meson spectra at $\theta=0$. 
At nonzero values of $\theta$, the mass gap of the system is expected to become smaller and some quantum numbers that distinguish mesons are no longer exact, so that we give their several improvements in section~\ref{sec:strategy_for_theta} to find the precise results in such cases. 

\begin{figure}[htb]
\centering
\includegraphics[scale=0.5]{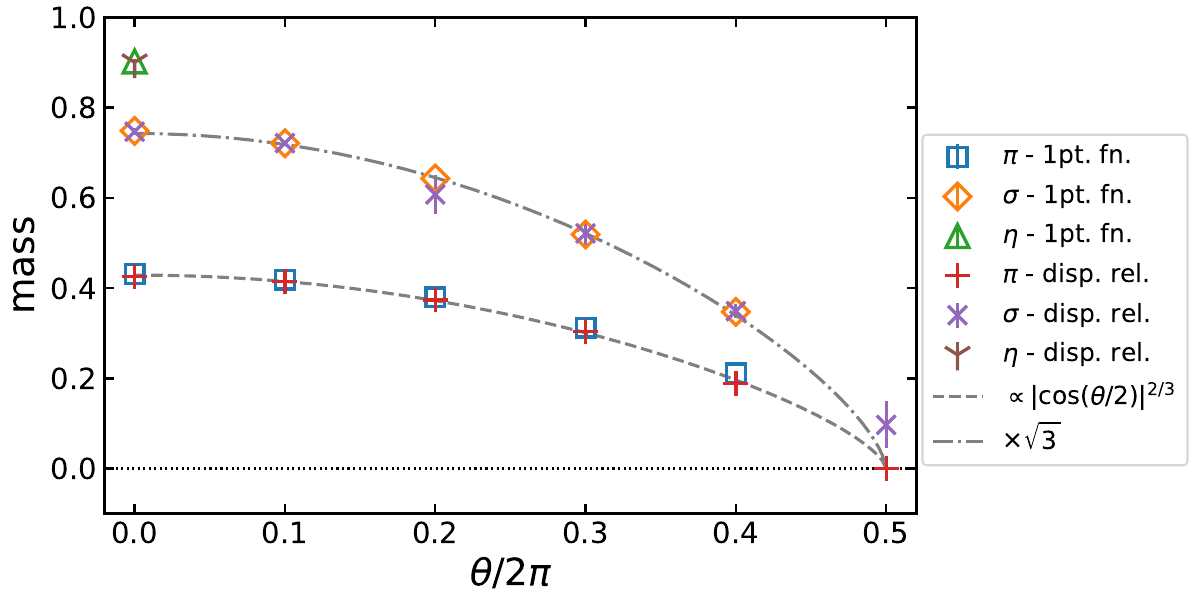}
\caption{\label{fig:mass_comp}
The $\theta$-dependent meson masses obtained by the one-point-function and dispersion-relation schemes are compared.
The gray dashed curve denotes the analytic calculation in the bosonized model, 
$M_{\pi}(\theta)=M_{\pi}(0)|\cos(\theta/2)|^{2/3}$, where the overall coefficient $M_{\pi}(0)$ 
is determined by averaging the results of the two schemes at $\theta=0$.
The gray dash-dot curve denotes the predicted sigma-meson mass, $\sqrt{3}M_{\pi}(0)|\cos(\theta/2)|^{2/3}$.}
\end{figure}

Now, we summarize the $\theta$-dependent masses of the pion and sigma meson in figure~\ref{fig:mass_comp}, 
which are obtained by the improved one-point-function and dispersion-relation schemes.
The eta-meson mass at $\theta=0$ is also plotted for reference.
As we can see in the plot, the results of the two schemes are consistent with each other.
We also compare the numerical results with the analytic calculation 
in the bosonized model shown in section~\ref{subsec:spectrum_bosonization}.
The analytic result of the pion mass (\ref{eq:M_pi_theta}) is depicted by the gray dashed curve 
in figure~\ref{fig:mass_comp}, normalized by the numerical data at $\theta=0$.
We find that the numerical result agrees with the analytic prediction.
Furthermore, the sigma-meson mass satisfies the relation, $M_{\sigma}/M_{\pi}=\sqrt{3}$,
even for large $\theta$ as shown by the gray dash-dot curve.
This result indicates that the semiclassical analysis by the WKB-type approximation
gives almost the correct answer for the wide range of $\theta$.
It is still interesting to clarify the mechanism of this agreement and the range of applicability.

At $\theta=\pi$, it is expected that the $2$-flavor Schwinger model becomes almost CFT-like due to the exponentially small mass gap.
In this case, it is not feasible to obtain the meson masses from the one-point functions unless we can manage the exponentially large system size.
Instead, we compare the one-point functions themselves with the calculation in the WZW model on a finite interval.
As shown in section~\ref{sec:Result_CFT}, the numerical results agree with the analytic calculation in the bulk, 
which provides numerical evidence that the $2$-flavor Schwinger model at $\theta=\pi$ is well approximated by the $SU(2)_1$ WZW model.
In addition to the WZW behavior, we find the additional constant contribution to the one-point function.
This may give the hint of the spontaneous breaking of the $(\mathbb{Z}_2)_{G+\mathrm{chiral}}$ symmetry, 
which requires more systematic comparisons of various volumes and also the continuum limit to be conclusive.

The eta meson is no longer the stable particle for $\theta\neq 0$ since the $G$-parity is broken.
Indeed, the behaviors of the one-point and correlation functions of the eta meson are quantitatively different from the other mesons.
We consider it consistent with the theoretical expectations for the $\eta \rightarrow \pi \pi $ decay and the $\eta$-$\sigma$ mixing.

Overall, our results are consistent with the analytic predictions for the massive $2$-flavor Schwinger model at small mass with nonzero $\theta$. 
We would like to emphasize that our calculation covers the region with the large $\theta$, 
which could not be accessed by the reweighting technique in the conventional Monte Carlo method~\cite{Fukaya:2003ph}. 
This becomes possible thanks to the efficiency of the DMRG for the lattice Hamiltonian formalism. 
It would be interesting and important to extend the applicability of the Hamiltonian-based approach to other systems, 
that suffer from the sign problem in the Monte Carlo methods, such as the finite-density systems in higher dimensions.

\acknowledgments
We would like to thank S.~Aoki, M.~Honda, K.~Murakami, S.~Takeda, and A.~Ueda for their useful discussions.
The numerical calculations were carried out on XC40 at YITP in Kyoto University and the PC clusters at RIKEN iTHEMS.
The work of E.~I. was partially supported by
the JSPS
(Grants No. 
JP21H05190, 
JP23H05439), 
and JST PRESTO Grant Number JPMJPR2113. 
This work was partially supported by Japan Society for the Promotion of Science (JSPS) KAKENHI Grant No. 23K22489 (Y.T.), and by Center for Gravitational Physics and Quantum Information (CGPQI) at Yukawa Institute for Theoretical
Physics.

\appendix

\section{Compact boson on the finite interval}
\label{sec:compact_boson_interval}

In this appendix, we analytically study the behaviors of various one-point functions 
of the $2$-flavor Schwinger model at $\theta=\pi$ on the open interval. 
The low-energy properties of the $2$-flavor Schwinger model at $\theta=\pi$ 
are well described by the $\SU(2)$ level-$1$ Wess-Zumino-Witten (WZW) model 
with the marginally relevant $J\bar{J}$ deformation. 
When the size of the interval is not so large, the $J\bar{J}$ deformation is negligible 
and we should be able to use the $\SU(2)_1$ WZW CFT as an effective description. 
This is equivalent to the compact boson at the self-dual radius, 
which is a free theory, so we can compute various correlation functions analytically. 

Within the approximations discussed above, the effective action becomes 
\begin{equation}
    \mathcal{L}_{\mathrm{eff}}=\frac{1}{4\pi}(\partial_\mu\varphi)^2, 
\end{equation}
with $\varphi(x)\sim \varphi(x)+2\pi$. 
The mapping of operators between the Schwinger model and the effective theory is given by 
\begin{align}
    &\overline{\psi}\psi \leftrightarrow -\cos\frac{\theta}{2}\cos\varphi, \quad 
    -\overline{\psi}i\gamma_5\psi \leftrightarrow \sin\frac{\theta}{2}\cos\varphi, \\
    &\overline{\psi}\tau_3\psi \leftrightarrow -\sin\frac{\theta}{2}\sin\varphi, \quad 
    -\overline{\psi}i\gamma_5\tau_3\psi \leftrightarrow -\cos\frac{\theta}{2}\sin\varphi. 
\end{align}
For the open interval, we need to specify the boundary condition, 
and we here use the Dirichlet boundary condition, $\varphi(x_1=0,x_2)=\varphi(x_1=L,x_2)=0$. 
Let us then compute 
\begin{equation}
    \langle \mathrm{e}^{\pm i\varphi(x_*)}\rangle=\frac{1}{Z}\int \mathcal{D}\varphi\exp(-\int \frac{1}{4\pi}(\partial \varphi)^2 d^2 x \pm i \varphi(x_*)). 
    \label{eq:pathint_1ptfunc}
\end{equation}
As we have the symmetry $\varphi\to -\varphi$ including the boundary condition, 
we have $\langle \mathrm{e}^{i\varphi(x_*)}\rangle = \langle \mathrm{e}^{-i\varphi(x_*)}\rangle$ so let us focus on the $+$ sign. 
The classical equation of motion is 
\begin{equation}
    \frac{1}{2\pi}\partial^2 \varphi(x)+i \delta(x-x_*)=0, 
\end{equation}
and we denote its solution as $\varphi_{\mathrm{cl}}$. 
Shifting $\varphi$ in \eqref{eq:pathint_1ptfunc} as $\varphi_{\mathrm{cl}}+\varphi$, we obtain 
\begin{align}
    \langle e^{i\varphi(x_*)}\rangle
    &=\frac{1}{Z}\int \mathcal{D} \varphi \exp\left(-\int \frac{1}{4\pi}(\partial \varphi+\partial\varphi_{\mathrm{cl}})^2 d^2 x +i(\varphi(x_*)+\varphi_{\mathrm{cl}}(x_*) )\right) \nonumber\\
    &=\mathrm{e}^{-\int \frac{1}{4\pi}(\partial \varphi_{\mathrm{cl}})^2 + i \varphi_{\mathrm{cl}}(x_*)}\nonumber\\
    &=\mathrm{e}^{i \varphi_{\mathrm{cl}}(x_*)/2}. 
\end{align}
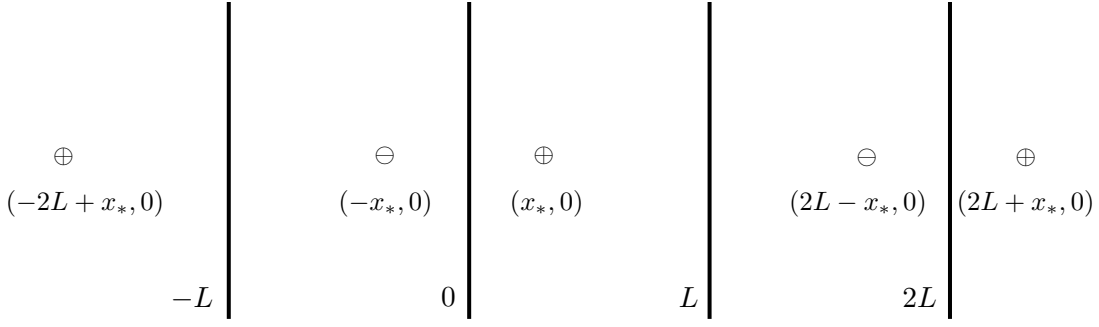
\begin{figure}[tb]
\centering
\tikzset{every picture/.style={line width=0.75pt}} 

\begin{tikzpicture}[x=0.75pt,y=0.75pt,yscale=-1,xscale=1]

\draw [line width=1.5]    (149.8,70.33) -- (149.8,229.73) ;
\draw [line width=1.5]    (269.8,70.33) -- (269.8,229.73) ;
\draw [line width=1.5]    (389.8,70.33) -- (389.8,229.73) ;
\draw [line width=1.5]    (509.8,70.33) -- (509.8,229.73) ;

\draw (254.33,212.4) node [anchor=north west][inner sep=0.75pt]    {$0$};
\draw (119.33,212.4) node [anchor=north west][inner sep=0.75pt]    {$-L$};
\draw (373,212.4) node [anchor=north west][inner sep=0.75pt]    {$L$};
\draw (484.67,212.4) node [anchor=north west][inner sep=0.75pt]    {$2L$};
\draw (221,141.73) node [anchor=north west][inner sep=0.75pt]  [font=\small]  {$\ominus $};
\draw (300.33,141.73) node [anchor=north west][inner sep=0.75pt]  [font=\small]  {$\oplus $};
\draw (461.67,142.4) node [anchor=north west][inner sep=0.75pt]  [font=\small]  {$\ominus $};
\draw (541,142.4) node [anchor=north west][inner sep=0.75pt]  [font=\small]  {$\oplus $};
\draw (60.67,142.07) node [anchor=north west][inner sep=0.75pt]  [font=\small]  {$\oplus $};
\draw (288.67,163.4) node [anchor=north west][inner sep=0.75pt]  [font=\small]  {$( x_{*} ,0)$};
\draw (203,163.4) node [anchor=north west][inner sep=0.75pt]  [font=\small]  {$( -x_{*} ,0)$};
\draw (428,163.4) node [anchor=north west][inner sep=0.75pt]  [font=\small]  {$( 2L-x_{*} ,0)$};
\draw (511.67,163.4) node [anchor=north west][inner sep=0.75pt]  [font=\small]  {$( 2L+x_{*} ,0)$};
\draw (37,163.4) node [anchor=north west][inner sep=0.75pt]  [font=\small]  {$( -2L+x_{*} ,0)$};

\end{tikzpicture}
\caption{The schematic picture of the setup for the mirror-image method.}
\label{fig:mirrorimage}
\end{figure}
Thus, the problem reduces to find the classical solution $\varphi_{\mathrm{cl}}$ satisfying the boundary condition, 
and we can use the mirror-image method by extending the domain of $\varphi_{\mathrm{cl}}$ (see figure~\ref{fig:mirrorimage}): 
\begin{align}
    \frac{1}{2\pi}\partial^2 \varphi_{\mathrm{cl}}&=-i \sum_{n=-\infty}^{\infty}\left[\delta(x-(x_*+2nL))-\delta(x-(-\bar{x}_*+2nL))\right]\nonumber\\
    &=-\frac{i}{2\pi}\partial^2 \sum_n \log \frac{|x-x_*+2nL|}{|x+\bar{x}_*+2nL|}, 
\end{align}
where we use the complex coordinate. Thus, 
\begin{equation}
    \varphi_{\mathrm{cl}}(x)=-i \sum_n  \log \frac{|x-x_*+2nL|}{|x+\bar{x}_*+2nL|}=-i \log \left|\frac{\sin\left(\frac{\pi}{2L}(x-x_*)\right)}{\sin\left(\frac{\pi}{2L}(x+\bar{x}_*)\right)}\right|. 
\end{equation}
Substituting this result formally, we find 
\begin{equation}
    \langle e^{i\varphi(x_*)}\rangle=\mathrm{e}^{i \varphi_{\mathrm{cl}}(x_*)/2}= \left(\frac{\sin \frac{\pi \epsilon}{2L}}{\sin\left(\frac{\pi}{2L}(x_*+\bar{x}_*)\right)}\right)^{1/2},
\end{equation}
where we introduce the parameter $\epsilon$ for the point-splitting regularization. 
With the normal ordering procedure at the scale $\rho$, 
we end up with multiplying the factor $\exp(\frac{1}{2}K_0(\rho \epsilon))\simeq \sqrt{\frac{2}{e^{\gamma}\rho \epsilon}}$ 
as $\epsilon\to 0$, and we get 
\begin{equation}
    \langle N_\rho[e^{i \varphi(x_*)}]\rangle = \sqrt{\frac{\pi}{e^\gamma \rho L}}\frac{1}{\sqrt{\sin \left(\frac{\pi}{2L}(x_*+\bar{x}_*) \right)}}. 
\end{equation}
Setting $x_*$ to be real (i.e. the location of the source is set at $x_2=0$), we have 
\begin{equation}
    \langle N_\rho[\cos \varphi(x_*)]\rangle = \sqrt{\frac{\pi}{e^\gamma \rho L}} \frac{1}{\sqrt{\sin (\pi x_*/L)}}, \quad \langle N_\rho[\sin \varphi(x_*)]\rangle =0. 
\end{equation}

Therefore, for example, the expectation value of $\overline{\psi}i\gamma_5 \psi$ behaves as follows 
with the Dirichlet boundary condition, when assuming the $SU(2)_1$ WZW CFT as the effective description at $\theta=\pi$: 
\begin{align}
    \langle \overline{\psi}i \gamma_5 \psi (x_*)\rangle 
    &= - 2C \sqrt{\mu\rho} \langle N_\rho[\cos(\varphi(x_*))]\rangle \notag\\
    &= - \sqrt{ \frac{e^\gamma \mu}{\pi L} } \frac{1}{\sqrt{\sin (\pi x_*/L)}} . 
\end{align}
On the other hand, we get $\langle \overline{\psi}\tau_3\psi\rangle=0$ in this flavor-symmetric Dirichlet boundary condition.

\section{Definition of the observables}
\label{sec:operators}

In this appendix, we show the explicit forms of the observables used to obtain the mass spectrum.
Let us start with the lattice version of the scalar and pseudo-scalar operators, 
$S_{f,n}$ and $PS_{f,n}$, for the flavor $f$ at the site $n$.
Their explicit forms are given by rewriting the local operators $\bar{\psi}\psi$ and $-i\bar{\psi}\gamma^5\psi$ 
in terms of the staggered fermions~\eqref{eq:fermion_dictionary}.
If we choose the same basis of the $(1+1)$d gamma matrices as in section~\ref{subsec:lattice_Hamiltonian}, 
we obtain 
\begin{equation}
S_{f,n}:=\frac{1}{4a}(-1)^{n}(-\chi_{f,n-1}^{\dagger}\chi_{f,n-1}+2\chi_{f,n}^{\dagger}\chi_{f,n}-\chi_{f,n+1}^{\dagger}\chi_{f,n+1}),
\label{eq:S_staggered}
\end{equation}
\begin{equation}
PS_{f,n}:=\frac{i}{4a}(-1)^{n}(\chi_{f,n-1}^{\dagger}\chi_{f,n}-\chi_{f,n}^{\dagger}\chi_{f,n-1}
-\chi_{f,n}^{\dagger}\chi_{f,n+1}+\chi_{f,n+1}^{\dagger}\chi_{f,n}),
\label{eq:PS_staggered}
\end{equation}
for the lattice sites $n=1,2,\cdots,N-2$.
Here we take the average of the three neighboring sites to remove the oscillation from the staggered phase.

Next, we define some global observables to measure 
the isospin quantum numbers $J_{z}$, $\bm{J}^{2}$, and the total momentum $K$.
These operators are used in the dispersion-relation scheme.
Unlike the local operators defined above, these global operators act on the whole lattice.
Let us consider the isospin operators (\ref{eq:J_cont}). 
Using the staggered fermion, 
the lattice version of $J_{z}$ in terms of the staggered fermions is given by 
\begin{equation}
J_{z}=\frac{1}{2}\sum_{n=0}^{N-1}\left(\chi_{1,n}^{\dagger}\chi_{1,n}-\chi_{2,n}^{\dagger}\chi_{2,n}\right).
\label{eq:J_z}
\end{equation}
It is convenient to define the lattice version of $J_{\pm}=J_{x}\pm iJ_{y}$ by 
\begin{equation}
J_{+}=\sum_{n=0}^{N-1}\chi_{1,n}^{\dagger}\chi_{2,n},
\label{eq:J_plus}
\end{equation}
\begin{equation}
J_{-}=\sum_{n=0}^{N-1}\chi_{2,n}^{\dagger}\chi_{1,n}.
\label{eq:J_minus}
\end{equation}
Then the isospin Casimir operator $\bm{J}^{2}$ can be defined 
as the combination of the operators above by 
\begin{equation}
\bm{J}^{2}=\frac{1}{2}(J_{+}J_{-}+J_{-}J_{+})+J_{z}^{2}.
\label{eq:J2}
\end{equation}
We can evaluate the expectation values of these operators systematically 
using the so-called matrix product operator.

Finally, we define the total momentum operator, 
which can be used to identify the momentum excitation~\cite{Banuls:2013jaa}. 
The continuum description of the gauge invariant momentum operator is given by 
\begin{equation}
K=\sum_{f=1}^{N_{f}}\int dx\,\psi_{f}^{\dagger}(i\partial_{x}-A_{1})\psi_{f},
\end{equation}
which commutes with the Hamiltonian of the continuum theory 
under the periodic boundary condition using the Gauss-law constraint.
In our case with the open boundary condition, $K$ does not commute with the Hamiltonian, 
so it is no longer the exact quantum number.
However, the momentum operator is still useful as an approximation to obtain the dispersion relation. 
After the gauge fixing $U_{n}=1$, the lattice version of the momentum operator is defined by 
\begin{equation}
K=\frac{i}{4a}\sum_{f=1}^{N_{f}}\sum_{n=1}^{N-2}(\chi_{f,n-1}^{\dagger}\chi_{f,n+1}-\chi_{f,n+1}^{\dagger}\chi_{f,n-1}).
\label{eq:K_staggered}
\end{equation}
This operator does not exactly commute with the terms $H_{w}$ (\ref{eq:H_w}) and $H_{J}$ (\ref{eq:H_J}) 
of the lattice Hamiltonian due to the open boundary and the finite lattice spacing effect.

\section{Results of the correlation-function scheme}
\label{sec:cf_scheme}

In the Euclidean lattice gauge theory, 
the mass spectrum is obtained from the correlation function in the imaginary time direction.
In our previous work focusing on $\theta=0$, 
we consider an analogous approach in the Hamiltonian formalism, 
examining the equal-time spatial correlation function. 
In this appendix, we extend this approach, the correlation-function scheme, 
to the case of $\theta\neq 0$ and show the results for the stable mesons, namely the pion and sigma meson.
The correlation function of the unstable eta meson is discussed in appendix~\ref{subsec:cf_eta}.
As we mentioned in section~\ref{sec:def-mixing-matrix}, the two-point connected correlation functions, 
\begin{equation}
C_{\pi}(r)=\Braket{\pi(x)\pi(y)}_{c},
\end{equation}
\begin{equation}
C_{\sigma}(r)=\Braket{\sigma(x)\sigma(y)}_{c},
\end{equation}
with the distance $r=|x-y|$ are obtained as the eigenvalues of the correlation matrix \eqref{eq:C_mat}. 
To reproduce the correct asymptotic behavior of the correlation functions, 
the cutoff parameter $\varepsilon$ has to be small enough in DMRG, 
resulting in a large bond dimension of the MPS.

\begin{figure}[htb]
\centering
\includegraphics[scale=0.45]{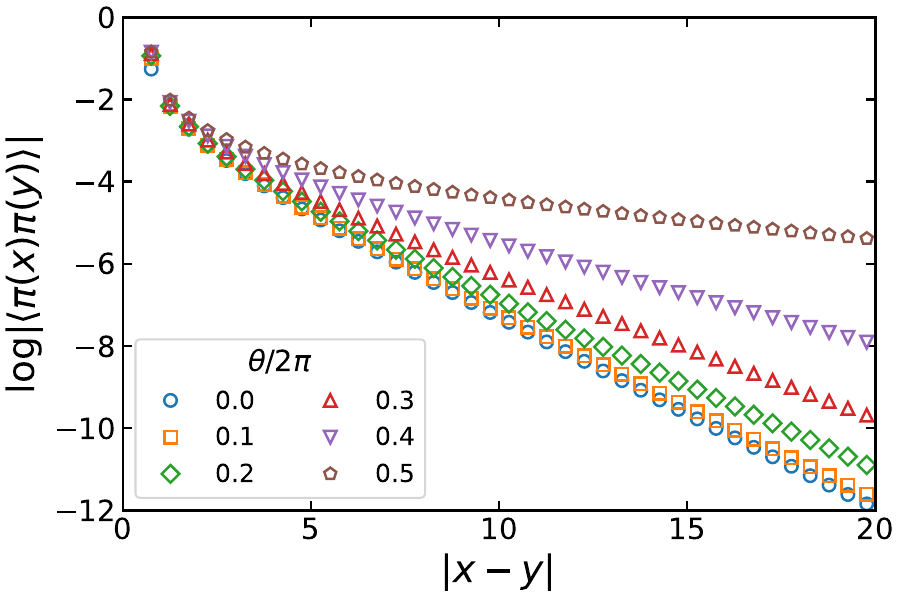}
\includegraphics[scale=0.45]{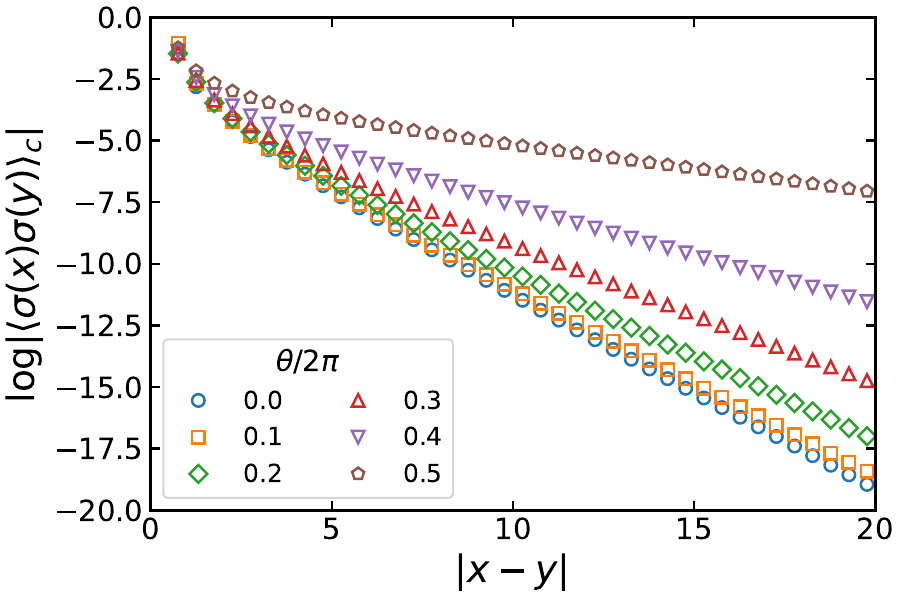}
\caption{\label{fig:cf_mesons}
The correlation function of the pion $\log|\Braket{\pi(x)\pi(y)}_{c}|$ (left) 
and the sigma meson $\log|\Braket{\sigma(x)\sigma(y)}_{c}|$ (right) 
are plotted against the distance $r=|x-y|$. 
The cutoff parameter in DMRG is $\varepsilon=10^{-16}$.}
\end{figure}

We generated the MPS of the ground state for $0 \leq \theta \leq \pi$ 
with $\varepsilon=10^{-10}$, $10^{-12}$, $10^{-14}$, and $10^{-16}$.
The number of lattice sites is set to $N=160$ in this scheme.
Then the correlation functions of the pion and sigma meson are computed 
by changing $x$ and $y$ symmetrically as $x=(L-r)/2$ and $y=(L+r)/2$ for $0 \leq r \leq L/2$. 
The results with $\varepsilon=10^{-16}$ are shown in figure~\ref{fig:cf_mesons} in log scale. 
The change of the gradient reflects the change of the mass depending on $\theta$.

To determine the mass of each meson, we use the so-called effective mass, 
namely the logarithmic derivative of the correlation function $C_{\mathcal{O}}(r)$, 
\begin{equation}
M_{\mathcal{O},\mathrm{eff}}(r)=-\frac{d}{dr}\log C_{\mathcal{O}}(r).
\end{equation}
On the lattice, we define the effective mass by the three-point average, 
\begin{equation}
M_{\mathcal{O},\mathrm{eff}}(r):=\frac{1}{4}\tilde{M}_{\mathcal{O},\mathrm{eff}}(r-2a)+\frac{1}{2}\tilde{M}_{\mathcal{O},\mathrm{eff}}(r)+\frac{1}{4}\tilde{M}_{\mathcal{O},\mathrm{eff}}(r+2a),
\label{eq:Meff_lat}
\end{equation}
where $\tilde{M}_{\mathcal{O},\mathrm{eff}}(r)$ 
is a discretized logarithmic derivative of the correlation function, 
\begin{equation}
\tilde{M}_{\mathcal{O},\mathrm{eff}}(r)=-\frac{1}{2a}\log\frac{C_{\mathcal{O}}(r+2a)}{C_{\mathcal{O}}(r)}.
\end{equation}
Here the factor $2a$ comes from the step size of changing $r$.

\begin{figure}[htb]
\centering
\includegraphics[scale=0.3]{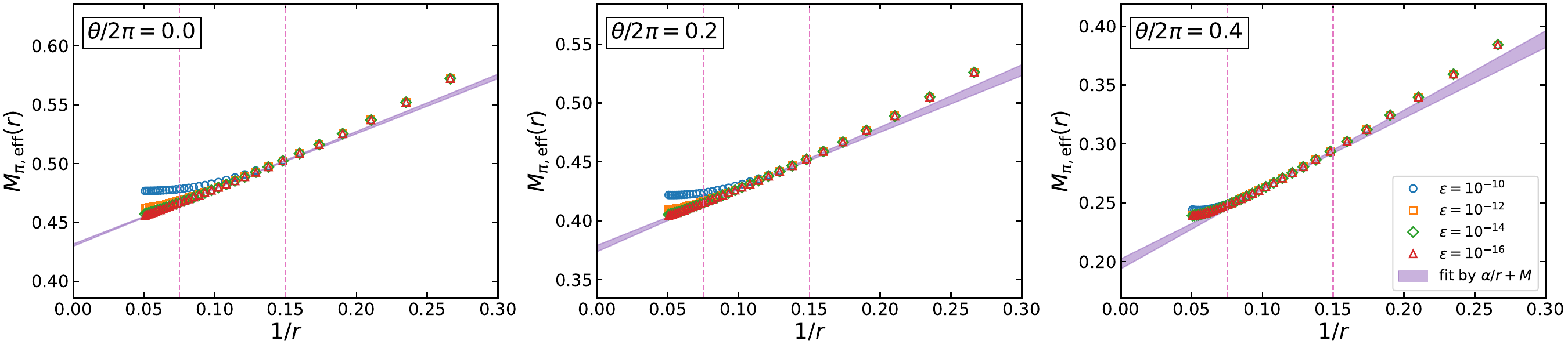}
\includegraphics[scale=0.3]{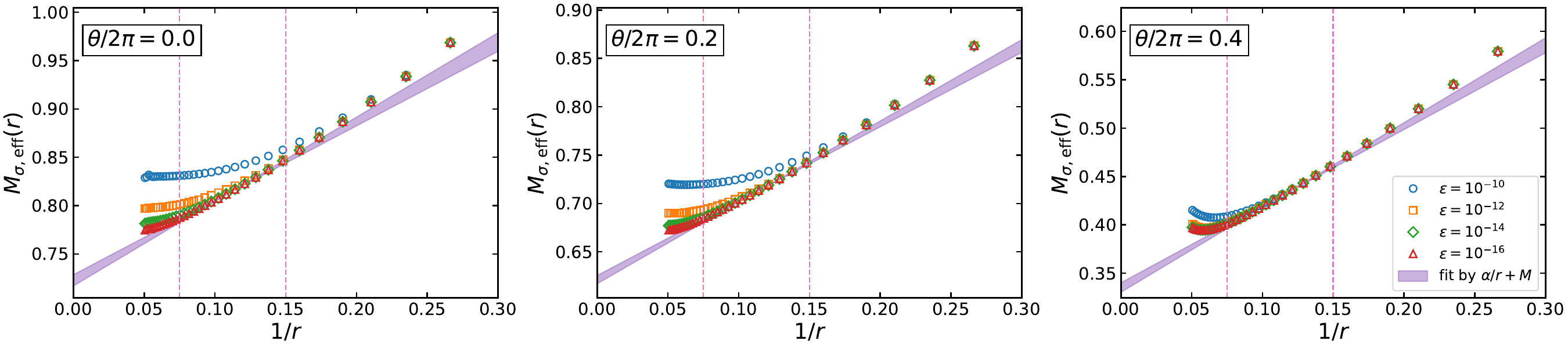}
\caption{\label{fig:Meff_pi_fit}
The effective masses $M_{\pi,\mathrm{eff}}(r)$ of the pion (top)
and $M_{\sigma,\mathrm{eff}}(r)$ of the sigma meson (bottom) 
are plotted against $1/r$ for various values of $\varepsilon$.
The left, center, and right columns are the results 
at $\theta/2\pi=0.0$, $0.2$, and $0.4$, respectively. 
The data points for $\varepsilon=10^{-16}$ are fitted by $\alpha/r+M$.
The fitting range is between the vertical dashed lines.
The fitting result is depicted by the shaded band with the systematic error.}
\end{figure}

Since we consider the spatial correlation function, 
the leading asymptotic behavior for $r\to\infty$ is not purely the exponential decay, 
but the Yukawa-type form, 
\begin{equation}
C_{\mathcal{O}}(r)\sim\frac{1}{r^{\alpha}}\exp(-Mr).
\label{eq:cf_yukawa}
\end{equation}
For example, we have $\alpha=1/2$ in the case of the $(1+1)$d free massive boson. 
The corresponding effective mass for the Yukawa-type correlator is given by 
\begin{equation}
M_{\mathcal{O},\mathrm{eff}}(r)\sim\frac{\alpha}{r}+M, 
\label{eq:Meff_asym}
\end{equation}
where an $O(1/r)$ contribution exists on top of the actual meson mass $M$. 
With this in mind, let us see our numerical results.
The effective masses computed from the correlation function of the pion and sigma meson 
are plotted against $1/r$ in the top and bottom rows of figure~\ref{fig:Meff_pi_fit}, respectively.
As $\varepsilon$ is decreased, namely the bond dimension is increased, 
the effective mass approaches $\propto 1/r$ asymptotically. 
This behavior is observed for $\theta\neq 0$ as well as for $\theta=0$, 
although the results are affected by the boundary when $r$ approaches the system size $L$.

Assuming the asymptotic behavior of Yukawa type (\ref{eq:Meff_asym}), 
we estimate the meson mass by the linear extrapolation $1/r\rightarrow 0$ 
of the effective mass (\ref{eq:Meff_lat}),
which is performed by fitting the data points by $\alpha/r+M$. 
Here the MPS generated with $\varepsilon=10^{-16}$ is used 
to suppress the effect of truncating the bound dimension as much as possible.
To evaluate the systematic errors from the uncertainty of the fitting range, 
we try fitting many times by changing the fitting range inside a given maximum region, 
and we obtain a histogram of the fitting results. 
The best-fitting result and its error are estimated from the position and width of the peak.
We set the maximum fitting region $0.075 \leq 1/r \leq 0.15$.
The fitting results for the pion and sigma meson are summarized 
in table~\ref{tab:fit_cf_pi} and~\ref{tab:fit_cf_sigma}, respectively.
The fitting curves are depicted by the purple shadows in figure~\ref{fig:Meff_pi_fit}.

\begin{table}[htb]
\centering
\begin{tabular}{|c|c|c|}\hline $\theta/2\pi$ & $M_{\pi}$ & $\alpha$ \tabularnewline\hline \hline 0.0 & 0.431(1) & 0.476(9) \tabularnewline\hline 0.1 & 0.416(1) & 0.49(1) \tabularnewline\hline 0.2 & 0.376(3) & 0.51(2) \tabularnewline\hline 0.3 & 0.306(4) & 0.55(4) \tabularnewline\hline 0.4 & 0.201(4) & 0.62(4) \tabularnewline\hline 0.5 & 0.031(4) & 0.88(4) \tabularnewline\hline \end{tabular}
\vspace{-1.3em}
\caption{\label{tab:fit_cf_pi}
The fitting results of the effective mass $M_{\pi,\mathrm{eff}}(r)$ of the pion by $\alpha/r+M$.
The errors of these values come from the systematic error from the uncertainty of the fitting range.}
\end{table}

\begin{table}[htb]
\centering
\begin{tabular}{|c|c|c|}\hline $\theta/2\pi$ & $M_{\sigma}$ & $\alpha$ \tabularnewline\hline \hline 0.0 & 0.722(6) & 0.83(5) \tabularnewline\hline 0.1 & 0.699(4) & 0.79(4) \tabularnewline\hline 0.2 & 0.624(4) & 0.78(4) \tabularnewline\hline 0.3 & 0.503(3) & 0.81(3) \tabularnewline\hline 0.4 & 0.334(5) & 0.85(4) \tabularnewline\hline 0.5 & 0.134(6) & 0.64(5) \tabularnewline\hline \end{tabular}
\vspace{-1.3em}
\caption{\label{tab:fit_cf_sigma} 
The fitting results of the effective mass $M_{\sigma,\mathrm{eff}}(r)$ of the sigma meson by $\alpha/r+M$.
The errors of these values come from the systematic error from the uncertainty of the fitting range.}
\end{table}

\begin{figure}[htb]
\centering
\includegraphics[scale=0.5]{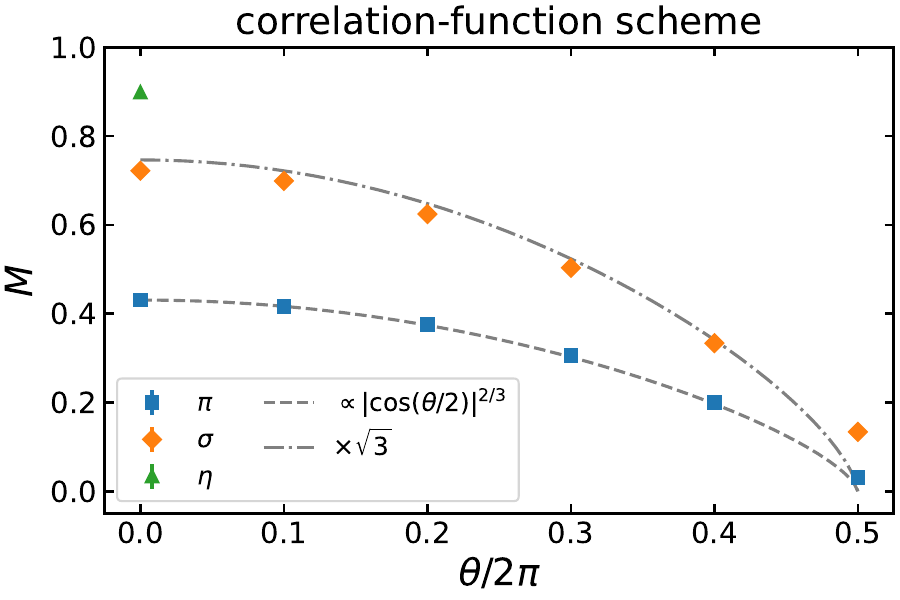} 
\caption{\label{fig:M_all_cf}
The masses of the pion and sigma meson obtained by the correlation-function scheme 
are plotted against $\theta/2\pi$. 
The eta-meson mass at $\theta=0$ is also plotted for reference.
The gray dashed curve denotes the analytic calculation by the bosonized model, 
$M_{\pi}(0)|\cos(\theta/2)|^{2/3}$, where the overall coefficient $M_{\pi}(0)$ 
is determined by the numerical result at $\theta=0$.
The gray dash-dot curve denotes $\sqrt{3}M_{\pi}(0)|\cos(\theta/2)|^{2/3}$.}
\end{figure}

The fitting of the effective mass at $\theta=\pi$ results in $M_{\pi} \sim 0$, 
which indicates the correlation function of the pion is 
almost power-law $C_{\pi}(r)\sim 1/r^{\alpha}$.
This result is reasonable since the system becomes CFT-like there.
On the other hand, we have a relatively large mass of the sigma meson 
at $\theta=\pi$ compared with the pion due to the boundary effect (finite size effect).
Indeed, the region where the $1/r$ behavior can be observed is less clear for the sigma meson.
The $\theta$-dependent meson masses are summarized in figure~\ref{fig:M_all_cf}.
The result is almost consistent with that of the one-point-function scheme in figure~\ref{fig:M_all_1pte} 
and the dispersion-relation scheme in figure~\ref{fig:M_all_disp}.
The $\theta$-dependence of the pion mass agrees with the calculation in the bosonized model in (\ref{eq:M_pi_theta}), 
and the sigma-meson mass is close to the result of the WKB approximation in (\ref{eq:sqrt3-Schwinger}).

\section{One-point-function scheme with the original setup}
\label{sec:1pt_func_original}

In section~\ref{subsec:improved_1pt_scheme}, we show the results 
of the improved one-point-function scheme on the lattice equipped with the wings.
In this appendix, we show the results of the one-point-function scheme 
with the naive open boundary condition used in our previous work~\cite{Itou:2023img}.
Then we check the consistency of these results.
Here the way of analysis is the same as section~\ref{subsec:improved_1pt_scheme}, 
and the difference is only the boundary condition.

\begin{figure}[htb]
\centering
\includegraphics[scale=0.45]{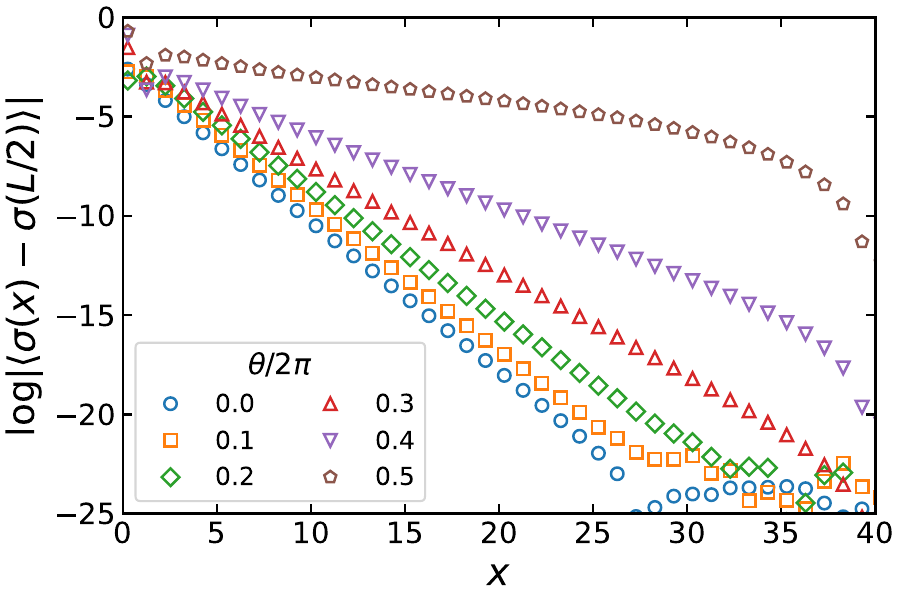}
\includegraphics[scale=0.45]{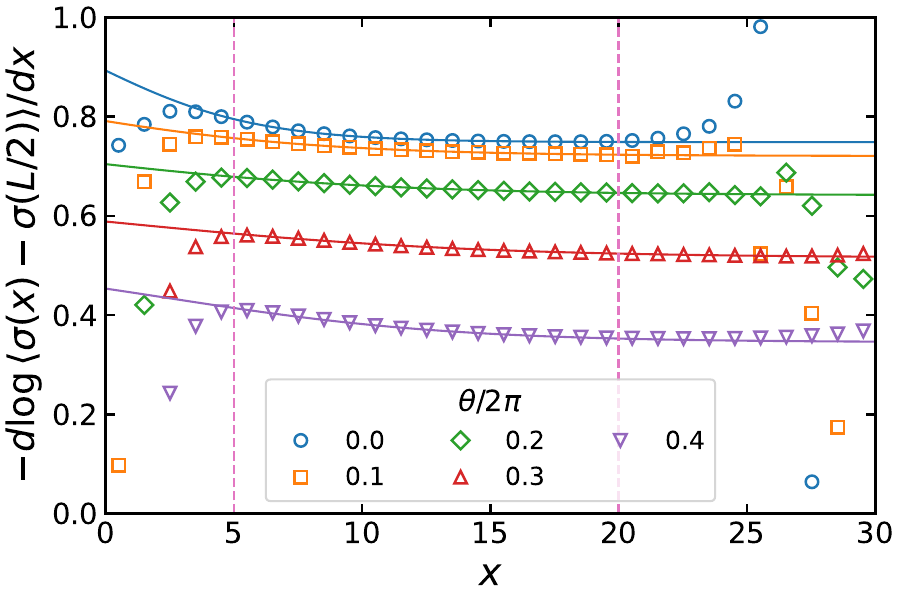} 
\caption{\label{fig:1pt_sigma}
(Left) The one-point function $\log|\Braket{\sigma(x)-\sigma(L/2)}|$ of the sigma meson 
is plotted against the distance $x$ from the boundary for $\varepsilon=10^{-10}$. 
The value at $x=L/2$ is subtracted from $\Braket{\sigma(x)}$ 
to eliminate the constant shift in the bulk. 
(Right) The effective mass calculated from the one-point function in the left panel 
is plotted against $x$. 
The results of fitting by (\ref{eq:Meff_1pt_ansatz}) are shown as well. 
The fitting range is between the vertical dashed lines.}
\end{figure}

The naive open boundary can be a source of the singlet meson.
We measure the one-point function $\Braket{\sigma(x)}$ of the sigma meson (\ref{eq:meson_op}) for the ground state, 
where we use the result in section~\ref{subsec:correlation_matrix} to determine $R_{+}$.
The result of $\Braket{\sigma(x)}$ is shown in the left panel of figure~\ref{fig:1pt_sigma}, 
which decays almost exponentially in the bulk region.
As we did in section~\ref{subsec:1pt_sigma}, 
we compute the effective mass (\ref{eq:Meff_1pt_lat}) 
and fit it by (\ref{eq:Meff_1pt_ansatz}) in the range $5\leq x\leq 20$. 
The result is shown in the right panel of figure~\ref{fig:1pt_sigma}, 
and the fitting parameters are summarized in table~\ref{tab:fit_1pt_sigma}.

\begin{table}[htb]
\centering
\begin{tabular}{|c|c|c|c|}\hline $\theta/2\pi$ & $M_{\sigma}$ & $\Delta M$ & $C$ \tabularnewline\hline \hline 0.0 & 0.74840(9) & 0.317(2) & 1.194(7) \tabularnewline\hline 0.1 & 0.7205(1) & 0.186(1) & 1.645(8) \tabularnewline\hline 0.2 & 0.6414(3) & 0.154(2) & 1.45(2) \tabularnewline\hline 0.3 & 0.5156(8) & 0.142(3) & 0.94(4) \tabularnewline\hline 0.4 & 0.3453(8) & 0.178(3) & 0.65(2) \tabularnewline\hline \end{tabular}
\vspace{-1.3em}
\caption{\label{tab:fit_1pt_sigma}
The fitting results of the effective mass computed 
from the one-point function $\Braket{\sigma(x)}$ of the sigma meson. 
The errors of these values come from the fitting error.}
\end{table}

\begin{figure}[htb]
\centering
\includegraphics[scale=0.45]{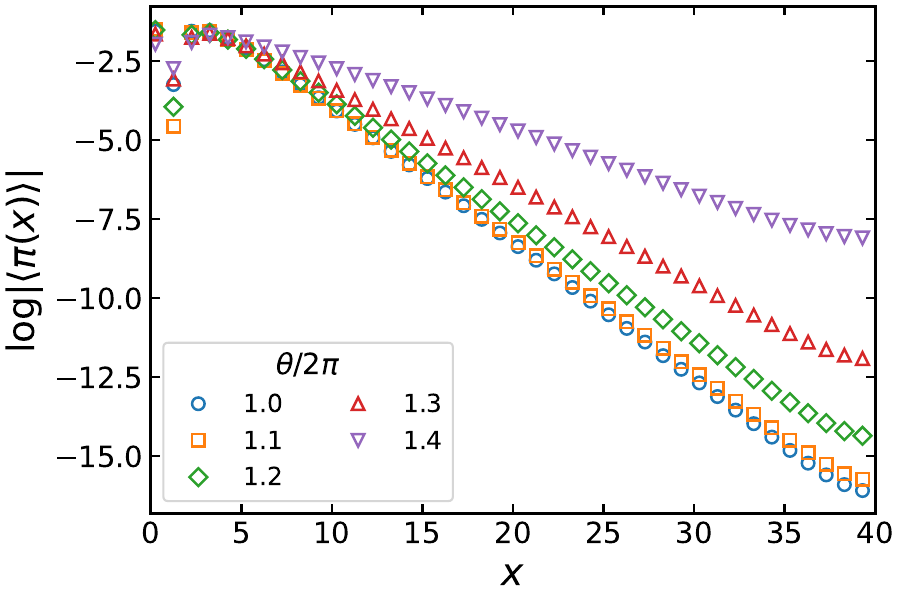}
\includegraphics[scale=0.45]{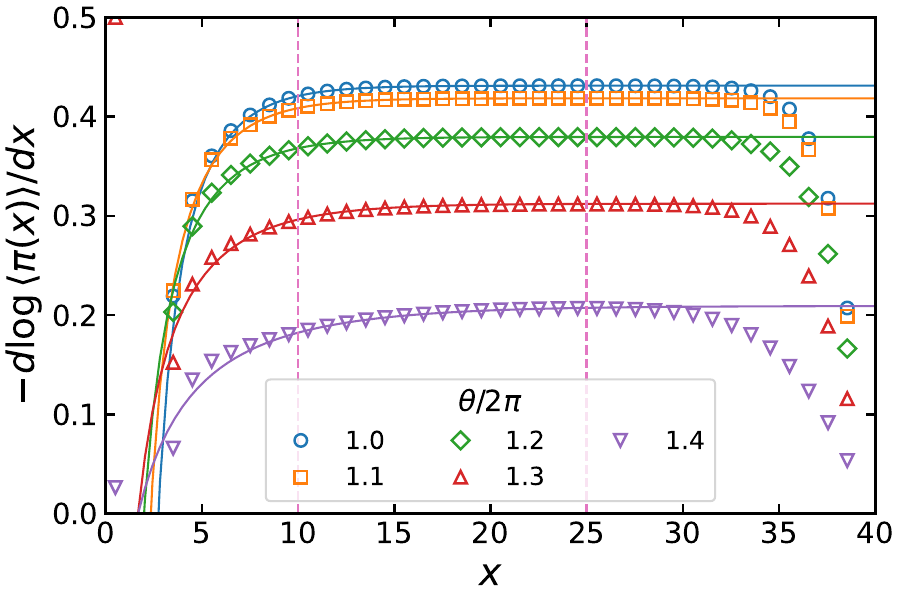} 
\caption{\label{fig:1pt_pi}
(Left) The one-point function $\log|\Braket{\pi(x)}|$ of the pion 
is plotted against the distance $x$ from the boundary for $\varepsilon=10^{-10}$. 
(Right) The effective mass calculated from the one-point function in the left panel is plotted against $x$. 
The results of fitting by (\ref{eq:Meff_1pt_ansatz}) are shown as well. 
The fitting range is between the vertical dashed lines.}
\end{figure}

In section~\ref{subsec:improved_1pt_scheme}, we applied the flavor-asymmetric chiral rotation 
to obtain the nonzero one-point function of the pion.
Here, we instead use the edge mode of the SPT state induced by shifting $\theta\rightarrow\theta+2\pi$ 
as a source of the pion, as we did in our previous work~\cite{Itou:2023img}.
We generate the ground states for $2\pi \leq \theta \leq 3\pi$ 
and measure the one-point function $\Braket{\pi(x)}$ of the pion (\ref{eq:meson_op}).
The result is shown in the left panel of figure~\ref{fig:1pt_pi}, 
which has similar behavior to the results with the twisted mass in figure~\ref{fig:1pte_pi}.
However, we find $\Braket{\pi(x)}\approx 0$ at $\theta = 3\pi$ unexpectedly.
The reason would be that the SPT state with the boundary charge 
gets higher energy than the trivial state due to the finite size effect 
around the critical point $\theta = 3\pi$.
Although we cannot compare the result at $\theta = 3\pi$ with the analytic calculation 
as in section~\ref{sec:Result_CFT}, we can still obtain the pion mass for $2\pi \leq \theta < 3\pi$.
We compute the effective mass (\ref{eq:Meff_1pt_lat}) 
and fit the results by (\ref{eq:Meff_1pt_ansatz}) in the range $10\leq x\leq 25$.
The results are plotted in the right panel of figure~\ref{fig:1pt_pi}, 
and the fitting parameters are summarized in table~\ref{tab:fit_1pt_pi}.

\begin{table}[htb]
\centering
\begin{tabular}{|c|c|c|c|}\hline $\theta/2\pi$ & $M_{\pi}$ & $\Delta M$ & $C$ \tabularnewline\hline \hline 1.0 & 0.431167(3) & 0.4178(6) & -0.631(3) \tabularnewline\hline 1.1 & 0.418404(2) & 0.3981(5) & -0.772(3) \tabularnewline\hline 1.2 & 0.379502(5) & 0.3501(8) & -0.960(6) \tabularnewline\hline 1.3 & 0.31225(1) & 0.272(1) & -1.187(9) \tabularnewline\hline 1.4 & 0.2092(1) & 0.167(2) & -1.36(1) \tabularnewline\hline \end{tabular}
\vspace{-1.3em}
\caption{\label{tab:fit_1pt_pi}
The fitting results of the effective mass computed from the one-point function $\Braket{\pi(x)}$ of the pion.
The errors of these values come from the fitting error.}
\end{table}

\begin{figure}[htb]
\centering
\includegraphics[scale=0.5]{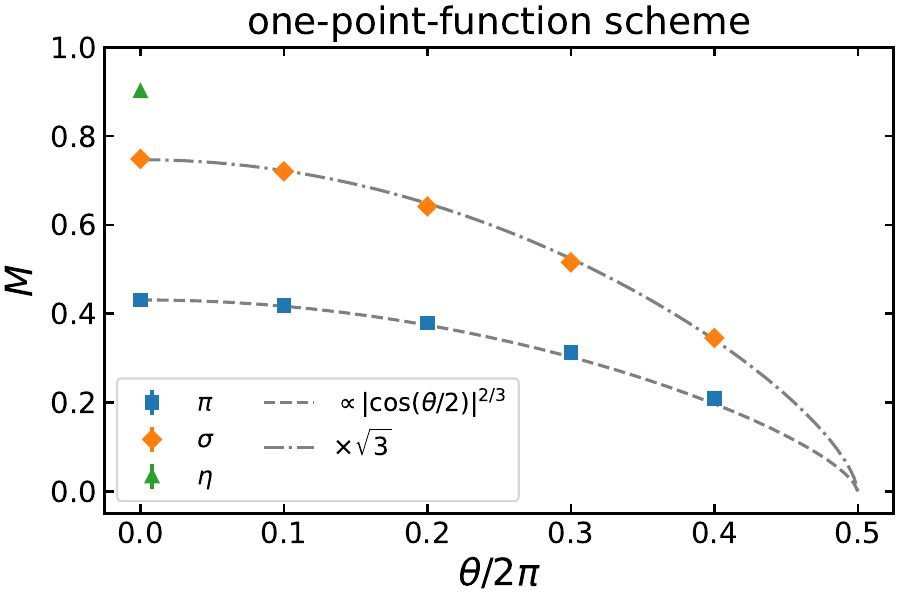} 
\caption{\label{fig:M_all_1pt}
The masses of the pion and sigma meson obtained by the one-point-function scheme 
with the naive open boundary are plotted against $\theta/2\pi$. 
The eta-meson mass at $\theta=0$ is also plotted for reference.
The gray dashed curve denotes the calculation by the bosonized model, 
$M_{\pi}(0)|\cos(\theta/2)|^{2/3}$, where the overall coefficient $M_{\pi}(0)$
is determined by the numerical result at $\theta=0$. 
The gray dash-dot curve denotes $\sqrt{3}M_{\pi}(0)|\cos(\theta/2)|^{2/3}$.}
\end{figure}

Let us summarize the results of the one-point-function scheme with the naive open boundary.
The $\theta$-dependent masses of the pion and sigma meson are shown in figure~\ref{fig:M_all_1pt}.
The results are consistent with those obtained on the lattice equipped with the wings in figure~\ref{fig:M_all_1pte}.
So, they also agree with the analytic results by the bosonization.
In the one-point-function scheme, we have many choices of the source of the mesons.
In this work, we tried four cases: the wings with the large fermion mass, 
the isospin-breaking chiral rotation, the naive open boundary, and the edge mode of the SPT state.
If we can find an appropriate source for the target particle, 
the one-point-function scheme should widely apply to other cases.

\section{Fate of the eta meson}
\label{sec:fate_of_eta}

At $\theta=0$, the eta meson is a stable particle protected by the $G$-parity.
However, the $G$-parity is explicitly broken for $\theta \neq 0$.
Then the eta meson is no longer stable but decays into the pions or mixes with the sigma meson.
In terms of the bosonized model, the eta meson has the interaction (\ref{eq:eta_interaction}) 
with the other mesons of the coupling $\sim\sin(\theta/2)$, 
as discussed in section~\ref{subsec:spectrum_bosonization}.
Indeed, the eigenstates of the Hamiltonian corresponding to the eta meson can be found at $\theta=0$, 
but such states disappear as we increase $\theta$ in the dispersion-relation scheme.
We observed that the eta-meson-like state and the sigma meson are mixed in the spectrum depending on $\theta$.
In this appendix, we show the one-point and two-point correlation functions of the eta meson 
and discuss how they behave when the eta meson is unstable.
We also show the values of the effective mass $M_\eta$ as a reference when we naively apply the estimation method discussed in the main text, and they should not be taken seriously as it is well-established only for stable particles. 

\subsection{One-point function of the eta meson}
\label{subsec:1pt_eta}

\begin{figure}[htb]
\centering
\includegraphics[scale=0.45]{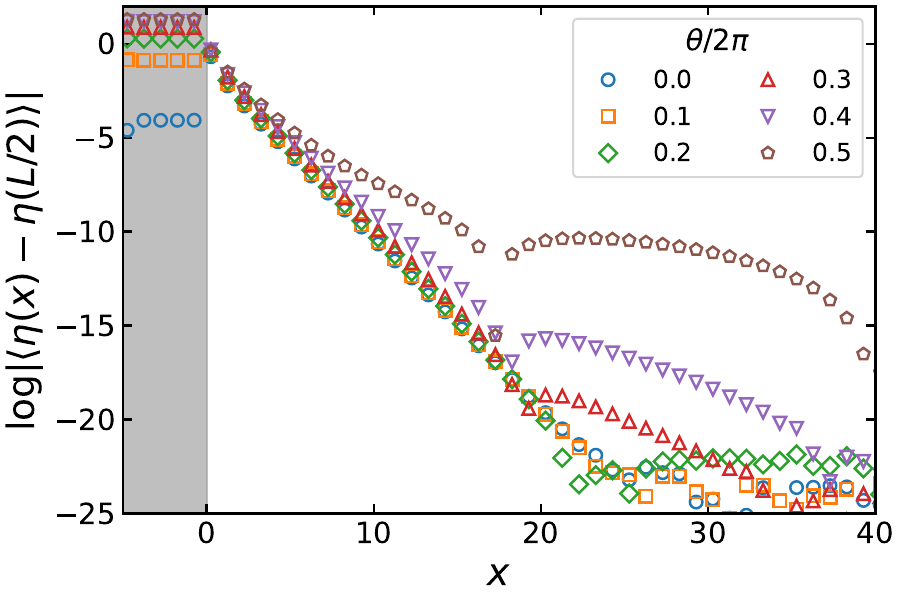}
\includegraphics[scale=0.45]{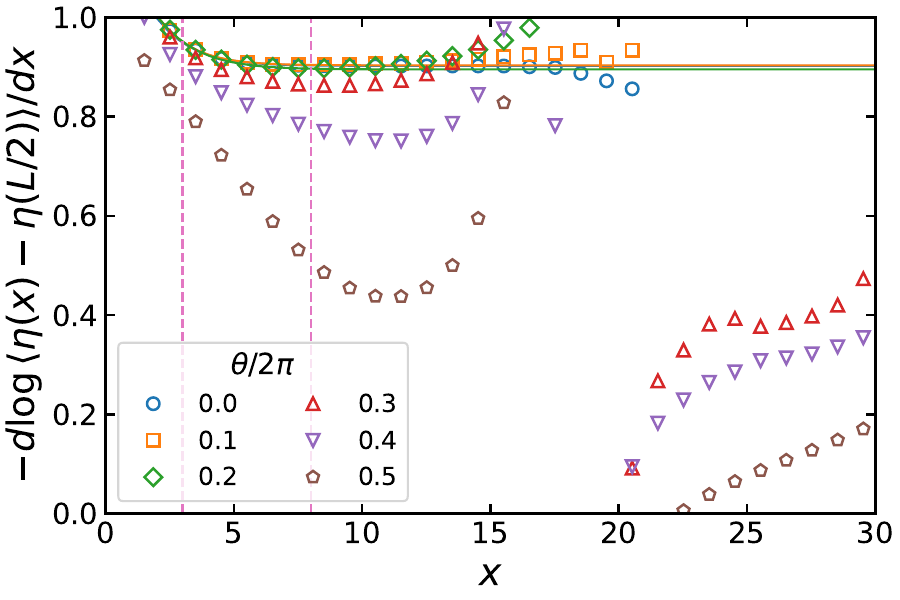} 
\caption{\label{fig:1pte_eta}
(Left) The one-point function $\log|\Braket{\eta(x)-\eta(L/2)}|$ 
of the eta meson is plotted against the distance $x$ from the boundary for $\varepsilon=10^{-10}$. 
The value at $x=L/2$ is subtracted from $\Braket{\eta(x)}$ to eliminate the constant shift in the bulk.
The shaded region indicates the wings.
(Right) The effective mass calculated from the one-point function 
in the left panel is plotted against $x$.
The results of fitting by (\ref{eq:Meff_1pt_ansatz}) are shown as well.
The fitting range is between the vertical dashed lines.}
\end{figure}

First, we look into the one-point function.
The boundary condition violating the $G$-parity can be a source of the eta meson.
Here we adopt the lattice setup with the wings region assigned a large fermion mass 
$m_{\mathrm{wings}}=m_0=10$ as in section~\ref{subsec:improved_1pt_scheme}.
We measure the one-point function $\Braket{\eta(x)}$ of the eta meson (\ref{eq:meson_op}) for the ground states, 
using the result in section~\ref{subsec:correlation_matrix} to determine the matrix $R_{+}$.
The results of $\Braket{\eta(x)}$ are shown in the left panel of figure~\ref{fig:1pte_eta}.
Unlike the pion and sigma meson, the behavior of $\Braket{\eta(x)}$ 
is no longer a simple exponential decay $\sim e^{-Mx}$ for $\theta\neq 0$, 
and it is not even monotonically decreasing resulting in the cusp of the plot.
The drastic change of the behavior depending on $\theta$ should be caused by the decay or mixing.
Note that the eta-meson mass $M_{\eta}\sim 0.9$ is larger than twice the pion mass $M_{\pi}\lesssim 0.43$,
and thus the decay is allowed in the current setup.

As long as the effect of decay is small, we can still measure 
the mass of the unstable eta meson by examining the one-point function.
Let us look into the effective mass.
The effective mass computed from $\Braket{\eta(x)}$ is shown 
in the right panel of figure~\ref{fig:1pte_eta}. 
As $\theta$ increases, the plateau region of the effective mass shrinks. 
We fit the effective mass in the short range, $3\leq x\leq 8$, 
by (\ref{eq:Meff_1pt_ansatz}) for $\theta/2\pi=0.0,0.1,0.2$.
The fitting results are plotted in the right panel of figure~\ref{fig:1pte_eta}, 
and the fitting parameters are summarized in table~\ref{tab:fit_1pte_eta}.
The $\theta$-dependence of the eta-meson mass at short distances is relatively small 
compared with the masses of the pion and sigma meson.
At long distances, $x\gtrsim 25$, 
the effective mass takes lower values of about 0.3 to 0.5 for $\theta/2\pi=0.3$ and $0.4$.
This behavior should be explained by the lighter states which the eta meson decays into.
It is interesting to check whether a similar behavior is observed in analytic calculations 
assuming the effective model of the decaying eta meson.

\begin{table}[htb]
\centering
\begin{tabular}{|c|c|c|c|}\hline $\theta/2\pi$ & $M_{\eta}$ & $\Delta M$ & $C$ \tabularnewline\hline \hline 0.0 & 0.9022(1) & 0.787(7) & 0.029(1) \tabularnewline\hline 0.1 & 0.90319(5) & 0.814(3) & 0.0239(5) \tabularnewline\hline 0.2 & 0.89501(8) & 0.731(3) & 0.0348(8) \tabularnewline\hline \end{tabular}
\vspace{-1.3em}
\caption{\label{tab:fit_1pte_eta}
The fitting results of the effective mass at short distances
computed from the one-point function $\Braket{\eta(x)}$ of the eta meson.
The errors of these values come from the fitting error.}
\end{table}

\subsection{Correlation function of the eta meson}
\label{subsec:cf_eta}

\begin{figure}[htb]
\centering
\includegraphics[scale=0.5]{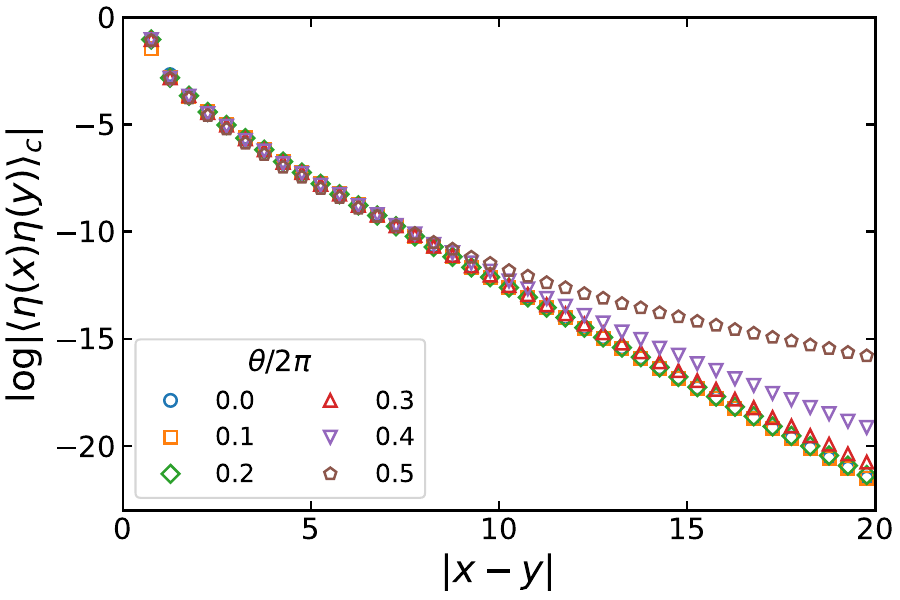} 
\caption{\label{fig:cf_eta}
The correlation function of the eta meson $\log|\Braket{\eta(x)\eta(y)}_{c}|$ 
is plotted against the distance $r=|x-y|$. 
The cutoff parameter in DMRG is $\varepsilon=10^{-16}$.}
\end{figure}

Next, let us focus on the spatial correlation function.
The connected correlation function of the eta meson, 
\begin{equation}
C_{\eta}(r)=\Braket{\eta(x)\eta(y)}_{c},
\end{equation}
is obtained as the eigenvalue of the correlation matrix (\ref{eq:C_mat}) 
together with the sigma meson.
The results are shown in figure~\ref{fig:cf_eta}.
In this plot, the behavior of the correlation function for $\theta/2\pi \gtrsim 0.4$ 
is qualitatively different from the other cases for the small $\theta$. 
The gradient of $\log\Braket{\eta(x)\eta(y)}_{c}$ suddenly changes at long distances $r \gtrsim 10$ 
whereas it is less sensitive to $\theta$ at short distances $r \lesssim 10$.

\begin{figure}[htb]
\centering
\includegraphics[scale=0.3]{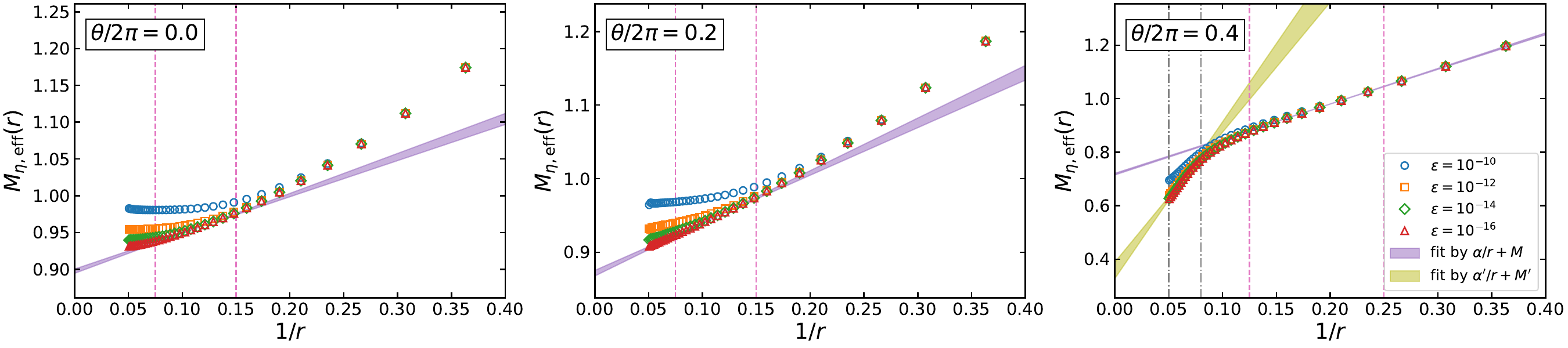}
\caption{\label{fig:Meff_eta_fit}
The effective masses $M_{\eta,\mathrm{eff}}(r)$ of the eta meson 
at $\theta/2\pi=0.0$ (left), $0.2$ (center), and $0.4$ (right) 
are plotted against $1/r$ for various values of $\varepsilon$. 
The data points for $\varepsilon=10^{-16}$ are fitted by $\alpha/r+M$. 
The fitting range is between the vertical dashed lines.
The fitting result is depicted by the shaded band with the systematic error.
For $\theta/2\pi=0.4$ we try two choices of the fitting range.
The result at the short distance is used to obtain the mass of the unstable eta meson.}
\end{figure}

Correspondingly, the behavior of the effective mass changes depending on the distance $r$.
The effective mass computed from the correlation function of the eta meson 
is shown in figure~\ref{fig:Meff_eta_fit} for $\theta/2\pi=0.0,0.2,0.4$.
For the large $\theta$ such as $\theta/2\pi=0.4$ in the right panel of figure~\ref{fig:Meff_eta_fit}, 
we find that $M_{\eta,\mathrm{eff}}(r)$ behaves as $\alpha/r+M$ 
at both short and long distances but with different $\alpha$ and $M$, 
which should result from the decay and mixing.
At short distances, the correlation function has information 
on the unstable eta meson before decaying. 
At long distances, the correlation function behaves like the lighter mesons, 
namely the pion after decaying or the mixed sigma meson.
From this point of view, we focus on the short-distance region 
to obtain the mass of the unstable eta meson when $\theta$ is large.
We fit the effective mass $M_{\eta,\mathrm{eff}}(r)$ by $\alpha/r+M$ for $\varepsilon=10^{-16}$.
Here we evaluate the systematic errors from the uncertainty of the fitting range 
by changing the range inside a maximum fitting region, as in appendix~\ref{sec:cf_scheme}.
The maximum region is $0.075 \leq 1/r \leq 0.15$ for $\theta/2\pi=0.0,\cdots,0.3$.
To see the short-distance behavior, the maximum region is changed 
to $0.125 \leq 1/r \leq 0.25$ for $\theta/2\pi=0.4$ 
and $0.15 \leq 1/r \leq 0.35$ for $\theta/2\pi=0.5$. 
The fitting results are depicted by the purple shadows in figure~\ref{fig:Meff_eta_fit}, 
and the results of the fitting parameter are summarized in table~\ref{tab:fit_cf_eta}.
The mass of the unstable eta meson is less sensitive to $\theta$ than the other mesons.

\begin{table}[htb]
\centering
\begin{tabular}{|c|c|c|}\hline $\theta/2\pi$ & $M_{\eta}$ & $\alpha$ \tabularnewline\hline \hline 0.0 & 0.899(3) & 0.51(3) \tabularnewline\hline 0.1 & 0.894(2) & 0.56(2) \tabularnewline\hline 0.2 & 0.872(4) & 0.68(4) \tabularnewline\hline 0.3 & 0.815(4) & 0.94(3) \tabularnewline\hline 0.4 & 0.720(3) & 1.30(2) \tabularnewline\hline 0.5 & 0.57(3) & 1.9(1) \tabularnewline\hline \end{tabular}
\vspace{-1.3em}
\caption{\label{tab:fit_cf_eta}
The fitting results of the effective mass $M_{\eta,\mathrm{eff}}(r)$ 
computed from the correlation function of the eta meson. 
The errors of these values come from the systematic error of fitting range.}
\end{table}

Next, we investigate the effective mass at long distances, 
where the contributions of the light mesons are expected to be significant.
To estimate the mass of the meson after decaying, 
we fit the effective mass $M_{\eta,\mathrm{eff}}(r)$ by $\alpha/r+M$ 
for $\theta/2\pi=0.4$ and $0.5$ with different fitting ranges than before.
For $\theta/2\pi=0.4$, we set the maximum fitting region $0.05 \leq 1/r \leq 0.08$ 
and obtain the results $M=0.34(3)$ and $\alpha=5.7(5)$.
The fitting result is depicted by the dark yellow shadow in the right panel of figure~\ref{fig:Meff_eta_fit}.
The obtained mass $M=0.34(3)$ is consistent with the sigma-meson mass at $\theta/2\pi=0.4$, 
for example, $M_{\sigma}=0.334(5)$ by the correlation-function scheme in table~\ref{tab:fit_cf_sigma}.
For $\theta/2\pi=0.5$, we set the maximum region $0.06 \leq 1/r \leq 0.125$ 
and obtain $M=-0.02(2)$ and $\alpha=6.1(3)$.
In this case, the mass is consistent with zero, 
which agrees with the CFT-like behavior of the pion and sigma meson at $\theta=\pi$.
It is again interesting to compare these results with analytic calculation assuming an effective model.

\bibliographystyle{JHEP}
\bibliography{Nf2_Schwinger,QFT}

\end{document}